\RequirePackage{fix-cm}

\documentclass[smallextended]{svjour3}       
\smartqed  

\usepackage{graphicx}
\usepackage{url}
\usepackage{booktabs}
\usepackage{latexsym}
\usepackage{multirow}
\usepackage{colortbl}
\usepackage{subcaption}
\usepackage[authoryear]{natbib}
\usepackage{adjustbox}
\usepackage{threeparttable}
\usepackage{textcomp}
\usepackage{xcolor}
\usepackage{stfloats}
\usepackage{bbding}
\usepackage{pifont}
\usepackage{hyperref}
\usepackage{makecell}
\usepackage{tabularx}
\usepackage{array}
\usepackage[most]{tcolorbox}
\usepackage{times}
\usepackage[T1]{fontenc}
\usepackage[utf8]{inputenc}
\usepackage{microtype}
\usepackage{inconsolata}
\usepackage[framemethod=TikZ]{mdframed}
\definecolor{mycolor}{RGB}{194, 214, 236}
\usepackage{amsmath,amssymb,amsfonts}
\usepackage[figuresright]{rotating}

\newcounter{finding}
\newcommand{\finding}[1]{\refstepcounter{finding}
 	\vspace{1mm}
	\begin{mdframed}[linecolor=gray,roundcorner=12pt,backgroundcolor=gray!15,linewidth=3pt,innerleftmargin=10pt,innertopmargin=6pt,innerbottommargin=6pt,leftmargin=0cm,rightmargin=0cm,topline=false,bottomline=false,rightline = false]
		\textbf{Findings \arabic{finding}:} #1
	\end{mdframed}
	\vspace{0.5mm}
}

\begin{document}

\title{An Empirical Study on the Effectiveness of Large Language Models for Binary Code Understanding}

\titlerunning{An Empirical Study on the Effectiveness of LLMs for Binary Code Understanding}       

\author{Xiuwei Shang \and Zhenkan Fu \and Shaoyin Cheng$^*$ \and 
        Guoqiang Chen$^*$ \and Gangyang Li \and 
        Li Hu \and Weiming Zhang \and Nenghai Yu
}


\institute{Xiuwei Shang, Zhenkan Fu, Gangyang Li and Li Hu \at
              University of Science and Technology of China, Hefei, China \\
              \email\{shangxw, ligangyang, pdxbshx\}@mail.ustc.edu.cn, buildxcbpro@gmail.com
            \and
           Guoqiang Chen \at
              QI-ANXIN Technology Research Institute, Beijing, China \\
              \email guoqiangchen@qianxin.com
           \and
           Shaoyin Cheng, Weiming Zhang and Nenghai Yu \at
              University of Science and Technology of China, Anhui Province Key Laboratory of Digital Security, Hefei, China \\
              \email\{sycheng, zhangwm, ynh\}@ustc.edu.cn 
}

\def\thefootnote{*}\footnotetext{Shaoyin Cheng and Guoqiang Chen are corresponding authors.}\def\thefootnote{\arabic{footnote}}

\date{Received: date / Accepted: date}

\maketitle

\begin{abstract}
Binary code analysis plays a pivotal role in the field of software security and is widely used in tasks such as software maintenance, malware detection, software vulnerability discovery, patch analysis, etc.
However, unlike source code, reverse engineers face significant challenges in understanding binary code due to the lack of intuitive semantic information. Although traditional reverse tools can convert binary code into C-like pseudo code, the lack of code comments and symbolic information such as function names still makes code understanding difficult.
In recent years, two groups of techniques have shown promising prospects: (1) Deep learning-based techniques have demonstrated competitive results in tasks related to binary code understanding, furthermore, (2) Large Language Models (LLMs) have been extensively pre-trained at the source-code level for tasks such as code understanding and generation. This has left participants wondering about the capabilities of LLMs in binary code understanding.
To this end, this work proposes a benchmark to evaluate the effectiveness of LLMs in real-world reverse engineering scenarios, which covers two key binary code understanding tasks, i.e., function name recovery and binary code summarization. To more comprehensively evaluate, we include binaries with multiple target architectures as well as different optimization options.
We gain valuable insights into the capabilities and limitations through extensive empirical studies of popular LLMs using our benchmark. Our evaluations reveal that existing LLMs can understand binary code to a certain extent, thereby improving the efficiency of binary code analysis. Our results highlight the great potential of the LLMs in advancing the field of binary code understanding, and provide new directions for binary code analysis techniques.

\end{abstract}

\section{Introduction}
\label{intro}

In the field of software security, binary code analysis plays a foundational role in tasks such as reverse engineering \citep{Canfora2011}, software vulnerability detection \citep{Giffin2004EfficientCI}, digital forensics \cite{naeem2022digital}, and patch analysis \cite{xu2017spain}, with engineers constantly dealing with vast amounts of unknown binary files. However, unlike human-readable source code, binary code that has been compiled, optimized, and stripped of symbol information \citep{Zhang2021sp} is like a "black box" devoid of semantic labels, where function names are simplified to placeholders, variable types are degraded to register operations, and code comments are completely absent. This semantic gap poses a huge challenge to reverse engineers in understanding binary code.

Although many decompilation tools, such as IDA Pro \citep{IDA}, Ghidra \citep{Ghidra} and BinaryNinja \citep{BinaryNinja}, can heuristically convert binary code into C-like pseudo code, alleviating some of the difficulties, they still lack easy-to-understand semantics information, especially function names and code comments that play an important role in comprehending the code \citep{Gellenbeck1991,nfre2021issta}. 
Recently, borrowing ideas from Natural Language Processing (NLP), deep learning-based methods have been proposed for understanding binary code. In the task of function name recovery, NERO \citep{nero2020}, NFRE \citep{nfre2021issta} and SymLM \citep{symlm2022ccs} utilized the disassembled assembly instruction sequence as neural models input to reassign descriptive names. NER \citep{Chen2023pst} utilized decompiled pseudo code with a higher abstraction level as input and achieves better performance.

Besides, the function name is only part of the semantic completion and is not enough to represent the complete behavioral logic of the code \citep{Sridhara2010ase}. If a natural language description can be generated for the binary code, it will greatly save the reverse engineer's analysis time. BinT5 \citep{bint52023saner} is the first pre-trained generative model designed specifically for binary code summarization, which is based on the source code model CodeT5 \citep{wang-etal-2021-codet5} and fine-tuned on binaries. Subsequently, as a unified multi-task pre-training model, HexT5 \citep{Xiong2023ase} can perform multiple downstream tasks such as code summarization and function name recovery. However, the expert methods mentioned above generally perform poorly when faced with unseen code samples, and their generalization capabilities still need to be improved.

Recently, breakthroughs in Large Language Models (LLMs) have brought new opportunities in this field. General LLMs, such as Llama \citep{Llama2023llama}, ChatGPT \citep{chatgpt2022training}, etc., have been widely demonstrated for their capabilities in natural language understanding and generation. Furthermore, LLMs in the programming language domain (e.g., CodeLlama \citep{codellama2023rozière} and WizardCoder \citep{wizardcoder2023luo}) have shown notable ability in program analysis tasks, like fixing security vulnerabilities \citep{HowEffective_ISSTA_2023}, test cases auto-generation \citep{zhang2023sectests}. These developments demonstrate the potential of LLMs to handle complex and structured information flows that are particularly important for understanding binary code. More strikingly, the few-shot learning property of LLMs enables them to quickly adapt to new domains via prompt engineering \cite{dai2022can}. This capability offers new possibilities for binary code analysis: Can LLMs bridge the representation gap between source code and binary code, and directly infer function semantics from decompiled pseudo code? Does the generated semantic information exhibit accuracy and interpretability comparable to that of professional reverse engineers? These questions have yet to be systematically answered in existing research.

In this paper, instead of developing a new technique, we investigate and compare the capabilities of various LLMs in understanding binary code. By harnessing the advanced semantic modeling and reasoning power of LLMs, we seek to explore the extent to which these models are able to understand binary code, a task that is traditionally handled by skilled human engineers \citep{David2020Neural}. To facilitate our evaluation, we design an automated approach to construct an evaluation benchmark dataset, which includes aligned source code, natural language summaries, and decompiled pseudo code. We then contrast the capabilities of LLMs on two binary code understanding tasks, namely: (1) function name recovery, and (2) binary code summarization. We extensively evaluated eight code domain LLMs (CodeGen \citep{codegen2_2023nijkamp}, WizardCoder \citep{wizardcoder2023luo}, DeepSeek-Coder \citep{deepseekcoder}, CodeLlama \citep{codellama2023rozière} et.al.), eight general domain LLMs (ChatGLM \citep{chatglm2023glm130b}, Vicuna \citep{vicuna2023judging}, Llama \citep{Llama2023llama}, Mistral \citep{mixtral2024jiang}, ChatGPT \citep{chatgpt2022training} et.al.), and four deep learning-based expert models (SymLM \citep{symlm2022ccs}, NER \citep{Chen2023pst}, BinT5 \citep{bint52023saner}, HexT5 \citep{Xiong2023ase}). Additionally, we explore the impact of injecting domain knowledge by fine-tuning LLMs on specific tasks. Furthermore, we conduct case studies in the context of virus analysis to showcase the performance of the LLMs in understanding binary code in real-world scenarios.

Our findings demonstrate that LLMs exhibit excellent potential in advancing automated binary code understanding. We call for more research in this area to further enhance the capabilities of LLMs to play a more critical role in the complex task of binary code analysis.

\noindent\textbf{Contributions.} In summary, the primary contributions of our work are as follows:

\begin{itemize}
    \item We design an automated method to construct a benchmark dataset to evaluate the capabilities of binary code understanding and release it to facilitate further research \footnote{\url{https://github.com/Sxxxw/BinaryLLMs-Eval}}.
    \item We conduct a thorough empirical study that evaluates the capabilities of eight code domain LLMs, eight general domain LLMs, and four DL-based methods on binary code understanding. Our primary focus lies on two fundamental tasks: function name recovery and binary code summarization.
    \item Our findings provide valuable insights into the capabilities and limitations of LLMs for understanding binary code. We thoroughly discuss the outcomes of our evaluations and offer suggestions for future research directions, aiming to propel advancements in this domain.
\end{itemize}

\noindent\textbf{Extended Version.} This paper is an extended version of our work published in the \emph{40th International Conference on Software Maintenance and Evolution} \citep{shang2024far}. Specifically, we extend the previous work in the following aspects:

\begin{itemize}
    \item The original study only targeted the x64 architecture and used the default optimization level of each project, without exploring the impact of different target architectures and optimization levels on binary code understanding. In extending this paper, we first substantially extend the evaluation analysis by assessing the effectiveness of LLMs in understanding binary code across four target architectures (\texttt{x86}, \texttt{x64}, \texttt{ARM}, \texttt{MIPS}) and four optimization options (\texttt{O0}, \texttt{O1}, \texttt{O2}, \texttt{O3}), respectively. (Corresponding to Section \ref{sec:rq1} and \ref{sec:rq2})
    \item We expand the scope of the experiment by incorporating additional LLMs for in-depth analysis, when investigating the key factors affecting the performance of LLMs in binary code understanding, as well as the impact of fine-tuning on performance. (Corresponding to Section \ref{sec:rq3} and \ref{sec:rq4})
    \item We conduct additional case studies to further demonstrate the practical role of LLMs in assisting reverse engineers in understanding binary code in real-world scenarios. (Corresponding to Section \ref{sec:rq5})
    \item We extend the scope of our discussion, particularly focusing on future work and limitations. (Corresponding to Section \ref{sec:discussions})
    \item We have also updated the related work with additional studies published in this research area, providing more detailed explanations of evaluation design, experimental metrics, etc.
\end{itemize}

\noindent\textbf{Structure of the Paper.} The rest of this paper is organized as follows: Section \ref{sec:background} summarizes the research background and related work, and explains our motivation. Detailed evaluation design and results analysis are presented in Section \ref{sec:design} and Section \ref{sec:results}, respectively. Subsequently, the discussion is thoroughly studied from multiple aspects in Section \ref{sec:discussions}. Finally, Section \ref{sec:conclusion} concludes this paper.

\begin{figure*}[t]
	\centering
        \scalebox{0.95}{
	\includegraphics[width=\linewidth]{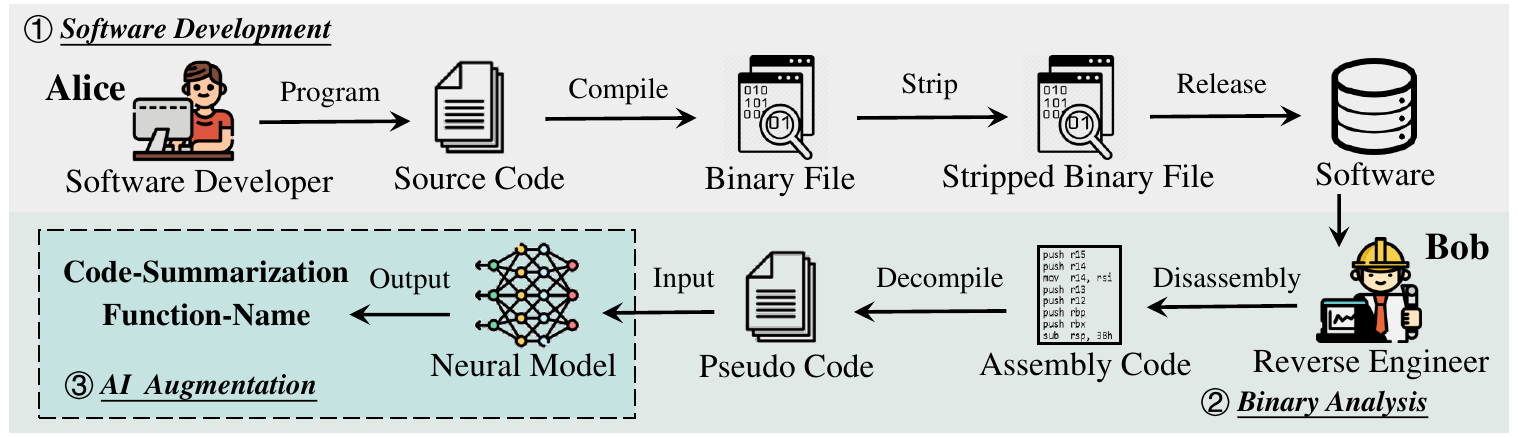}
        }
        \vspace{0ex}
	\caption{Application background of binary code understanding.}
    \vspace{-1ex}
    \label{fig:background}
\end{figure*}

\section{Background and Motivation \label{sec:background}}

\subsection{\textbf{Binary Code Understanding}}

\vspace{-3ex}

\begin{figure*}[h]
	\centering
        \scalebox{0.92}{
	\includegraphics[width=\linewidth]{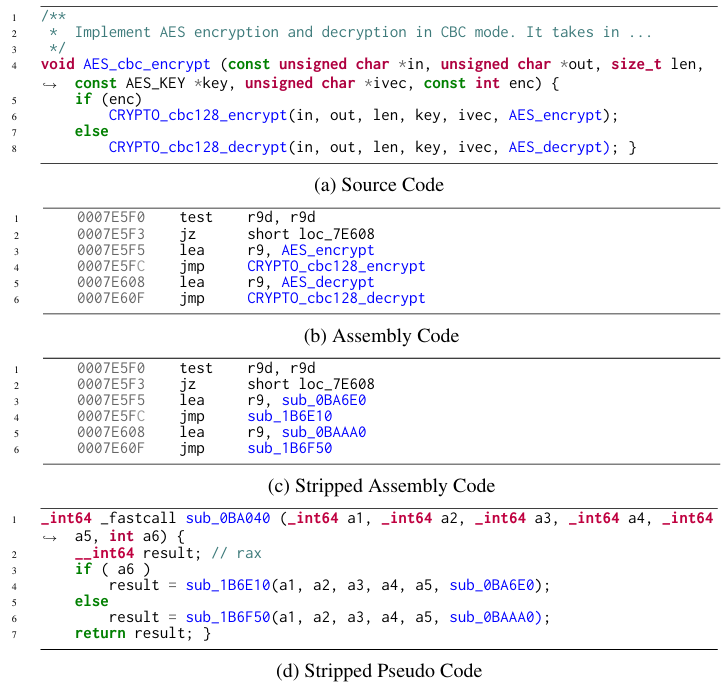}
        }
        \vspace{0ex}
    \caption{Example of source, assembly, stripped assembly and stripped pseudo code snippet.}
    \label{fig:code}
\end{figure*}

As shown in Figure \ref{fig:background}, consider a scenario where a software developer, Alice, who writes a program in a high-level language like C/C++. Her code, written in a human-readable format (Figure \ref{fig:code}a), must be translated into a form the computer can execute. This is where the compilation process comes into play, turning Alice's source code into binary code, which is a series of 1s and 0s that the machine can understand and execute.

Once compiled, Bob, a reverse engineer, wants to understand how Alice's program works. He uses a disassembler to convert the binary code back into an assembly code sequence (Figure \ref{fig:code}b), which is more readable than binary but still quite low-level and contains a large number of machine instructions. By doing this, Bob can get some basic structure of the program. However, Alice has used the “strip” command to remove symbol information from the original binary code in order to reduce file size and protect intellectual property. This makes Bob's job more difficult because he is now missing critical information such as function names (replaced with meaningless placeholders such as \texttt{sub\_0BA6E0}) and code comments (Figure \ref{fig:code}c).

Finally, Bob uses a decompiler in an attempt to reconstruct the high-level logic of the program. The decompiler generates pseudo code (Figure \ref{fig:code}d), an approximation of what Alice's original source code might have looked like. However, due to the complexity of the decompilation process and the lack of symbolic information in the stripped binary, the pseudo code may not exactly match Alice's original code, making it still difficult for Bob to understand the function and behavior of the program.

At this point, Bob attempts to use advanced natural language processing (NLP) techniques, such as LLMs or deep learning models, which are adept at identifying patterns and inferring logical structures. Bob leverages these techniques to \textbf{predict function names} and \textbf{generate natural language summaries} of the code's functionality. This process can be formalized as:
\begin{equation}
\small
\mathcal{N},\ \mathcal{S}=f(\mathcal{F}, \mathcal{B}) 
\end{equation}
where $\mathcal{F}$ is a stripped decompiled function in pseudo code form in binary file $\mathcal{B}$, which is fed into LLMs denoted as $f$. The objective is to generate a meaningful function name denoted as $\mathcal{N}$, and a natural language description denoted as $\mathcal{S}$ of this function.

Through this process, Bob combined the analytical power of AI with his reverse engineering skills to bridge the gaps left by the stripped binaries and gain a deeper understanding of Alice's original programming intent. Specifically, by recovering the function names and summarizing the code functions, Bob was able to quickly infer the role of each function, and then understand the design logic of the entire program in a relatively short period of time, which greatly improved the efficiency of reverse analysis.

\subsection{\textbf{Related Works}}

\subsubsection{\textbf{Function Name Recovery}}
The task of binary function name recovery aims to reassign descriptive function names to functions in binary files that have been stripped of symbolic information. During compilation and release, debugging symbols such as function names, variable names, and type information are often stripped to minimize file size, enhance security, and obscure implementation details. However, their absence complicates reverse engineering, security analysis, malware detection, and vulnerability discovery. Function name recovery helps researchers quickly extract critical function semantics and enhance the understanding of binary program behavior.

The research community has extensively explored the task of function name recovery. Initially, signature matching techniques \citep{ZaremskiW95} were applied to restore library function names. However, their adaptability to broader contexts posed challenges. As a result, probabilistic prediction approaches gained traction. A notable example is Debin \citep{Debin18ccs}, which leverages a conditional random field (CRF) model to infer debugging information. 

In recent years, the rapid progress of artificial intelligence technology has led to the widespread adoption of neural network-based methods for function name recovery. Among them, NERO \citep{nero2020} uses augmented control flow graphs, combined with the neural model of the encoder-decoder paradigm, to effectively capture the behavioral characteristics of functions and provide a new method for recovering binary function names. NFRE \citep{nfre2021issta} proposes a lightweight framework for function name recovery that utilizes the sequential and structural information of assembly instructions. The efficiency and scalability of the framework provide the possibility to process large-scale binary files. Based on NFRE, SymLM \citep{symlm2022ccs} further considers the calling context to help the model understand function semantics, and leverages advanced pre-training models \citep{pei2021trex} for instruction embedding. XFL \citep{sp2023} leverages feature engineering to extract constants and control flow, employs an aggregation strategy to integrate global and contextual embeddings. And utilizes PfastreXML \citep{JainPV16sigkdd} with binary function embeddings for efficient multi-label classification, addressing sparsity and class imbalance in function name labeling. NER \citep{Chen2023pst} starts from the perspective of binary code representation and studies the effectiveness of different representations for function name recovery using deep neural models, providing new perspectives and tools for this field. Finally, llasm \citep{llasm2025tosem} pioneers an encoder-decoder LLM fusion architecture for binary function name recovery, integrating a pre-trained assembly encoder with a natural language decoder. This approach leverages LLM reasoning to enhance semantic understanding, broaden function name prediction, and enable deeper binary code interpretation. These studies refine binary function name recovery through data representation, optimization effects, and NLP integration, advancing reverse engineering, malware analysis, and program comprehension.

\subsubsection{\textbf{Binary Code Summarization}}
Binary Code summarization aims to automatically describe the functionality of binary functions in natural language, assisting reverse engineers in analyzing binary files without source code. Due to the lack of high-level semantic information (such as function names and comments), decompiled pseudo-C code remains difficult to interpret. This task improves analysis efficiency through automatic summarization, facilitating malware detection and vulnerability discovery while aiding engineers in comprehending binary code behavior.

Recently, several approaches have emerged, each contributing unique solutions to the challenges of understanding and summarizing binary code. In these approaches, BinT5 \citep{bint52023saner} is the first model focused on binary code summarization, which extends the application scope of source code pre-trained language models. This model treats the decompiled code as a special programming language, uses fine-tuned CodeT5 \citep{wang-etal-2021-codet5} to capture its semantics and generate a summary. The introduction of BinT5 opens up new avenues for binary code summarization research. HexT5 \citep{Xiong2023ase} proposes a unified pre-training model also based on CodeT5, which allows multi-task learning, supports function name recovery, binary code summarization, and other downstream tasks, and demonstrates promising performance. CP-BCS \citep{YeW00D0JW23} proposes a framework based on control flow graphs and pseudo code for generating binary function summaries. This approach effectively captures the execution behavior of assembly code by combining bidirectional instruction-level control flow graphs and pseudo code, overcoming the challenges posed by the low-level representation of assembly code. Bin2Summary \citep{SongCZ24} enhances the semantic understanding of binary code fragments through function-specific code embedding techniques and utilizes an attention-based seq-to-seq model to generate natural language summaries from the embedded binary code. Lastly, MALSIGHT \citep{lu2024malsight} enhances binary code summarization by integrating reverse function extraction, recursive summarization, and static/dynamic annotations, capturing function call context to better handle malware's complex interactions. It fine-tunes an LLM on malware source code and benign pseudo code and introduces BLEURT-sum to improve summary accuracy and readability.

Additionally, multi-intent code summarization has become a research focus, aiming to generate customized summaries tailored to different developer needs and intents. MiSum \citep{zhu2025misum} introduces a multi-intent code summarization framework based on a multimodal heterogeneous code graph (MM-HCG), integrating assembly code (CFG) and pseudo code (AST) for multi-level code understanding. Utilizing an intent-aware attention mechanism, MiSum generates customized summaries tailored to different code analysis needs, enhancing both the flexibility of binary code summarization and its effectiveness in reverse engineering and cross-layer code analysis.

\subsection{\textbf{Large Language Models and Our Motivation}}
In recent years, Large Language Models (LLMs) have garnered widespread attention from both academia and industry due to their remarkable capabilities. Typically composed of billions or even trillions of parameters, LLMs are trained on vast amounts of data to learn the relationships between programming languages and natural language. Notable examples include GPT-3 \citep{gpt3} and LLaMA \citep{LLaMA}, all of which have demonstrated outstanding performance across various Natural Language Processing (NLP) tasks.  Amidst this surge in research interest, LLMs specifically designed for programming languages have rapidly emerged. These include Codex \citep{CodeX}, GPT-NeoX \citep{gpt-neox}, CodeT5+ \citep{codet5+}, PolyCoder \citep{ploycode}, WizardCoder \citep{wizardcoder2023luo}, and CodeLlama \citep{codellama2023rozière}, among others. These models have exhibited exceptional proficiency in code comprehension, further expanding the potential applications of LLMs in software development and analysis. Recently, A few studies \citep{HowEffective_ISSTA_2023,zhang2023sectests,hou2023large} find that LLMs have demonstrated excellent capabilities in dealing with natural language tasks, as well as source code understanding, indicating that they have the potential to be applied to complex analysis of source code. 

Traditional binary reverse engineering and decompilation tools, such as Ghidra \citep{Ghidra} and IDA Pro \citep{IDA}, play a crucial role in converting binary code into high-level languages. However, they exhibit significant limitations in terms of readability and comprehension of complex code structures. The lack of high-level semantic information and debugging symbols makes decompilation a labor-intensive process, heavily reliant on expert knowledge and domain-specific expertise.  Nevertheless, binary code shares inherent similarities with source code and natural language, as they all follow specific patterns and structures that can be learned and leveraged by LLMs \citep{Zhang2023lev}. 

Therefore, this study will explore the potential of LLMs in understanding binary code, aiming to evaluate whether these models can cross domain boundaries and extend their capabilities in natural language and source code to binary code analysis. This is expected not only to provide new perspectives for automated code understanding, but also to open up new application paths in areas such as reverse engineering and malware analysis.

\section{Evaluation Design \label{sec:design}}

\subsection{\textbf{Dataset Construction}}

\begin{figure*}[t]
	\centering
        \scalebox{1.0}{
	\includegraphics[width=\linewidth]{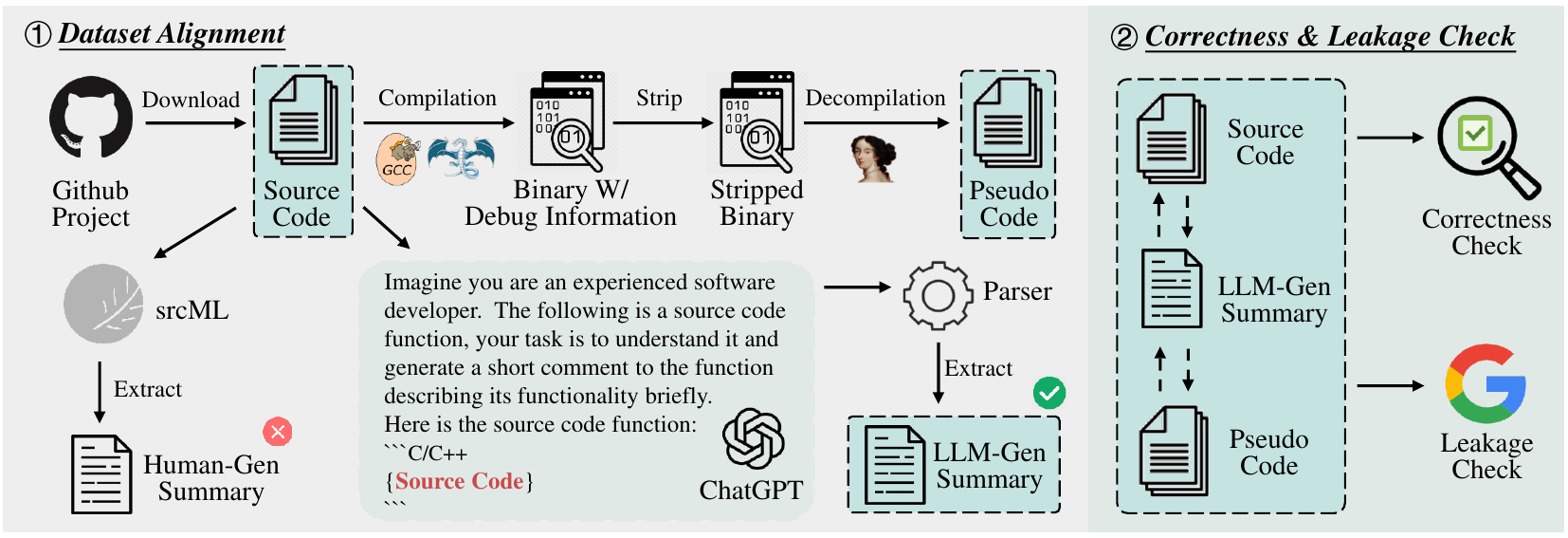}
         }
	\caption{An overview of the benchmark dataset construction process.}
    \vspace{-1ex}
    \label{fig:datacon}
\end{figure*}

\begin{sidewaystable*}[thp]
    \vspace{100ex}
    \caption{Statistics of our benchmark dataset.}
    \centering
    \renewcommand{\arraystretch}{1.20}
    \setlength{\tabcolsep}{1.3mm}
    \scalebox{0.99}{
    \begin{tabular}{lccrrrrrrrr}
        \toprule
            \multirow{2}{*}{\textbf{Project}} & \multirow{2}{*}{\textbf{Domain}} & \multirow{2}{*}{\textbf{\# Binaries}} & \multicolumn{7}{c}{\textbf{\# Aligned Functions}} & \multirow{2}{*}{\textbf{\# Select}} \\
            \cmidrule(r){4-10}
            & & & x64\_O0 & x64\_O1 & x64\_O2 & x64\_O3 & x86\_O0 & ARM\_O0 & MIPS\_O0 &  \\
        \midrule
            FFmpeg \citep{FFmpeg} & Video  & 14 & 44,157 & 26,215 & 22,722 & 21,295 & 45,443 & 42,979 & 44,518 & 250 \\
            Redis \citep{Redis} & Database & 14 & 9,688 & 8,827 & 7,736 & 7,072 & 9,716 & 5,713 & 5,486 & 200 \\
            Curl \citep{Curl} & Network & 14 & 1,329 & 995 & 796 & 706 & 1,364 & 1,329 & 1,329 & 200 \\
            Masscan \citep{Masscan} & Network & 7 & 946 & 674 & 524 & 456 & 946 & 946 & 946 & 150 \\
            Llama2.c \citep{Llama2.c} & Neural Network & 14 & 67 & 67 & 45 & 35 & 67 & 67 & 67 & 35 \\
            Whisper.cpp \citep{whisper.cpp} & Neural Network & 63 & 18,907 & 4,537 & 3,755 & 3,505 & 4,963 & 4,932 & 4,947 & 200 \\
            OpenSSL \citep{OpenSSL} & Crypto & 14 & 4,859 & 4,193 & 3,509 & 3,366 & 4,764 & 4,776 & 4,778 & 250 \\
            zstd \citep{zstd} & Compress & 7 & 2,302 & 956 & 694 & 557 & 2,284 & 2,279 & 2,279 & 200 \\
            ImageMagick \citep{ImageMagick} & Image & 21 & 7,112 & 4,207 & 3,737 & 3,631 & 6,560 & 6,560 & 6,560 & 200 \\
            Libvips \citep{Libvips} & Image & 7 & 3,721 & 3,147 & 2,860 & 2,747 & 4,233 & 3,724 & 3,723 & 208 \\
            Libexpat \citep{Libexpat} & Format & 7 & 260 & 234 & 155 & 134 & 260 & 260 & 260 & 100 \\
            Ultrajson \citep{Ultrajson} & Format & 14 & 27 & 11 & 8 & 7 & 26 & 26 & 26 & 7 \\
        \midrule
            \textbf{Total (12)} & 8 & 196 & 93,348 & 54,052 & 46,533 & 43,504 & 80,600 & 73,565 & 74,893 & 2,000 \\
        \bottomrule
    \end{tabular} }
    \vspace{0ex}
    \label{tab:dataset_construction} 
\end{sidewaystable*}

Before we can effectively evaluate the ability of LLMs to understand binary code, a comprehensive benchmark dataset is necessary to provide a consistent basis for different model evaluations and comparisons. The specific process of constructing the benchmark dataset is shown in Figure \ref{fig:datacon}. 

\subsubsection{\textbf{Source Code Selection}} 
\vspace{-1ex}
To reflect real-world reverse engineering needs, we believe that code sources of the benchmark should meet the following requirements:
\begin{itemize}
	\item \emph{Authenticity}: the code should come from real projects, not toy programs or incomplete code snippets. 
	
	\item \emph{Breadth}: the selected code should cover multiple fields and not be limited to specific fields or application scenarios. 
	
	\item \emph{High quality}: the selected code should be of good quality, including clear structure, reasonable naming conventions, etc. 

    \item \emph{Credibility}: the selected code should ideally be sourced from projects maintained by well-known or reputable developers or organizations to accurately reflect real-world application scenarios.
\end{itemize}

Therefore, as shown in Table~\ref{tab:dataset_construction}, we select 12 real-world projects implemented in C language with the highest star ratings on Github, including FFmpeg, Redis, Curl, Masscan, Llama2.c, Whisper.cpp, OpenSSL, zstd, ImageMagick, Libvips, Libexpat, and Ultrajson, which have high credibility, excellent code quality and maintenance standards, covering eight application domains, including crypto, compress, network, video, image, database, neural network, etc.

\subsubsection{\textbf{Compile, Strip and Decompile}} 
\vspace{-1ex}
We cross-compile these projects in different compilation environments to obtain binary files. Specifically, the target architectures include \texttt{x64}, \texttt{x86}, \texttt{ARM}, and \texttt{MIPS}, where the \texttt{x64} architecture uses four optimization levels (\texttt{O0} to \texttt{O3}) while the other architectures only use the \texttt{O0} optimization level.
As illustrated in Table~\ref{tab:dataset_construction}, we generate a total of 196 binaries. Subsequently, we employ the \texttt{strip} command in Linux to remove the symbol tables from these binaries to simulate binaries without symbol information that are common in actual reverse engineering.

Previous research \citep{Chen2023pst} has found that using pseudo code as a representation of binary code is more effective for neural models than assembly instruction sequence and Intermediate Representation (IR), as it provides a higher-level code representation that facilitates model understanding. Therefore, we directly utilize IDA Pro \citep{IDA} to decompile the binary files and convert the binary code into pseudo code form without considering other representation forms.

\subsubsection{\textbf{Alignment}}  
\vspace{-1ex}
We use DWARF \citep{Dwarfformat} debugging information to align source code and pseudo code, which can record functions, variables in binary functions, and their locations (include source file name, line number, and column number) in source code. As shown in Table~\ref{tab:dataset_construction}, we obtain a total of 466,495 functions matching source code and pseudo code. To align source code and human-written summary, we use srcML \citep{Maletic2015Exploration} to analyze and parse the source files, then collect single- and multi-line summaries above the location of function declarations and definitions. Through the above steps, the alignment of source code -- pseudo code -- human summary is finally achieved.

\subsubsection{\textbf{Ground-truth Identifiction}} 
\vspace{-1ex}
For the function name recovery task, we parse the function names in the source code as labels using regular expressions. For the binary code summarization task, we first consider using human-written comments extracted from source code files as labels as in previous work \citep{bint52023saner,Xiong2023ase}. However, we found that only about 14.8\% of functions have comments written by human developers. Worse yet, not all comments are describing the functional summary of the function, but will also contain some noisy content, and they are of varying quality and style. Therefore, using human-written comments as ground-truth is unreliable.

Presently, an increasing number of research works \citep{Dagdelen2024,Bzdok2024,tan2024large} employ large language models such as ChatGPT \citep{chatgpt2022training} for tasks like data annotation, and has demonstrated a certain degree of reliability. Inspired by these pioneering works, we utilize ChatGPT to generate summaries as ground-truth. Specifically, we use the source code of the function to construct the prompt shown in Figure \ref{fig:datacon}, prompting ChatGPT to generate a short summary describing the function's purpose and functionality.

\subsubsection{\textbf{Correctness \& Leakage Check}}
\vspace{-1ex}
It is crucial to ensure the correctness of the ground-truth, so we perform a correctness check on the descriptive summaries generated by ChatGPT. Specifically, we invited three experienced domain experts to review the match between the source code and the summary. Experts were asked to give each abstract a "pass" or "fail" score. If two or more experts give a "fail" rating, the data will be removed directly from the dataset; if one expert gives a "fail" rating, we will conduct a collective discussion and give a final in conclusion. Finally, as shown in Table~\ref{tab:dataset_construction}, we select 2,000 functions, each of which contains seven compilation settings (i.e., \texttt{x64\_O0}, \texttt{x64\_O1}, \texttt{x64\_O2}, \texttt{x64\_O3}, \texttt{x86\_O0}, \texttt{ARM\_O0}, \texttt{MIPS\_O0}), totaling 14,000 pieces of data as the benchmark dataset.

It is also imperative that benchmark datasets are not included in the training set of LLMs to mitigate the risk of data leakage. All our evaluation data are decompiled pseudo code, and the symbolic information is stripped away so that it is significantly different from the corresponding source code form, which greatly avoids data leakage. To further ensure the validity and reliability of our benchmark evaluation, we use the Google search engine to check whether the code appears on the Internet in clear text. The results show that none of the pseudo codes are retrieved by whole-word matching.

\begin{table*}[t]
  \caption{Detail information of Large Language Models we apply in this work. (In the License column, "\checkmark" indicates Open-source, "\texttimes" indicates Closed-source.)}
  \vspace{-2ex}
  \centering
  \renewcommand{\arraystretch}{1.1}
  \setlength{\tabcolsep}{0.55mm}{
  \scalebox{0.70}{
    \begin{tabular}{clcccccccc}
    \toprule
        \multirow{2}{*}{\textbf{Domain}} & \multicolumn{1}{c}{\multirow{2}{*}{\textbf{Model}}} & \multirow{2}{*}{\textbf{\makecell{Release\\Time}}}  & \multirow{2}{*}{\textbf{\makecell{Size}}} & \multirow{2}{*}{\textbf{\makecell{Base\\Model}}} & \multicolumn{3}{c}{\textbf{Training Corpus}} & \multirow{2}{*}{\textbf{Publisher}} & \multirow{2}{*}{\textbf{License}}  \\
        \cmidrule(r){6-8}
        & & & & & \textbf{Raw Size} & \textbf{\#Tokens} & \textbf{\#Instances} & \\
    \midrule
     \multicolumn{1}{c}{\multirow{8}{*}{\makecell{Code \\ LLMs}}} 
       & CodeGen25-7b-instruct \citep{codegen2_2023nijkamp} & Jul-2023 & 7B & CodeGen2 & - & 1.4T & - & Salesforce & \checkmark  \\
       & WizardCoder-15b-V1.0 \citep{wizardcoder2023luo} & Jun-2023 & 15B & StarCoder & - & - & 78.0K & WizardLM & \checkmark \\
       & WizardCoder-33b-V1.1 \citep{wizardcoder2023luo} & Jan-2024 & 33B & Deepseek-Coder & - & - & - & WizardLM & \checkmark \\
       & Code Llama-7b-instruct-hf \citep{codellama2023rozière} & Jun-2023 & 7B & Llama-2-7b & 4.4TB & 525.0B & - & Meta AI & \checkmark \\
       & Code Llama-13b-instruct-hf \citep{codellama2023rozière} & Jun-2023 & 13B & Llama-2-13b & 4.4TB & 525.0B & - & Meta AI & \checkmark  \\
       & Code Llama-34b-instruct-hf \citep{codellama2023rozière} & Jun-2023 & 34B & Llama-2-34b & 4.4TB & 525.0B & - & Meta AI & \checkmark  \\
       & Code Llama-70b-instruct-hf \citep{codellama2023rozière} & Jan-2024 & 70B & Llama-2-70b & - & 1.0T & - & Meta AI & \checkmark  \\
       & DeepSeek-Coder-33b-instruct \citep{deepseekcoder} & Nov-2023 & 33B & - & - & 2.0T & -  & DeepSeek-AI & \checkmark  \\
    \midrule
      \multicolumn{1}{c}{\multirow{7}{*}{\makecell{General \\ LLMs}}} 
       & ChatGLM2-6B \citep{chatglm2023glm130b} & Jun-2023 & 6B & - & - & 1.4T & - & THUDM & \checkmark \\
       & Vicuna-7b-v1.5 \citep{vicuna2023judging} & Aug-2023 & 7B & Llama-2-7b & - & - & 125.0K & L.Zheng et al. & \checkmark \\
       & Vicuna-13b-v1.5 \citep{vicuna2023judging} & Aug-2023 & 13B & Llama-2-13b & - & - & 125.0K & L.Zheng et al. & \checkmark \\
       & Llama-2-13b-chat-hf \citep{Llama2023llama} & Jul-2023 & 13B & - & - & 2.0T & - & Meta AI & \checkmark \\
       & Llama-2-70b-chat-hf \citep{Llama2023llama} & Jul-2023 & 70B & - & - & 2.0T & - & Meta AI & \checkmark  \\
       & Mistral-7B-Instruct-v0.2 \citep{mixtral2024jiang} & Dec-2023 & 7B & Mistral-7B & - & - & - & Mistral AI & \checkmark  \\
       & Mixtral-8x7B-Instruct-v0.1 \citep{mixtral2024jiang} & Dec-2023 & 47B & Mistral-7B & - & - & - & Mistral AI & \checkmark  \\
       & ChatGPT \citep{chatgpt2022training} & Nov-2022 & - & - & - & - & - & OpenAI & \texttimes \\
    \bottomrule
    \end{tabular} } }
    \vspace{-1ex}
  \label{tab:LLMs_info}%
\end{table*}

\subsection{\textbf{Large Language Models Setup}}
\vspace{-1ex}
\subsubsection{\textbf{Large Language Models As Is}}
\vspace{-1ex}
We select eight code domain LLMs, i.e., CodeGen25 \citep{codegen2_2023nijkamp}, DeepSeek-Coder \citep{deepseekcoder}, two versions of WizardCoder \citep{wizardcoder2023luo}, four versions of CodeLlama \citep{codellama2023rozière}, and select eight general domain LLMs, i.e., ChatGLM \citep{chatglm2023glm130b}, two versions of Vicuna \citep{vicuna2023judging}, two versions of Llama \citep{Llama2023llama}, two versions of Mistral \citep{mixtral2024jiang}, and ChatGPT \citep{chatgpt2022training}. The principles for our selection are: (1) state-of-the-art, (2) pre-trained on enough source code to be able to understand code to a certain extent, and (3) have text generation and code generation capabilities. In addition, in order to ensure that the model can follow the instructions, we all choose the instruct-tuned version. Table~\ref{tab:LLMs_info} provides detailed information, including parameter size, base model, training corpus, publisher, etc.

Limited by the context window length, we set the maximum input tokens to 4,096, and code snippets exceeding the length will be truncated. For the function name recovery task, we set the maximum new tokens to 48, and for the code summarization task, we set it to 256. We set the sampling temperature to 1, top\_p to 0.95, top\_k to 10, and num\_beams to 1. For all open-source models, we downloaded them from HuggingFace \citep{HuggingFace} and deployed on our local machine with FP16 mixed precision enabled during inference. For ChatGPT, we called its latest \texttt{gpt-3.5-turbo-16k} version through the OpenAI's API.

\begin{figure*}[t]
	\centering
        \scalebox{1.0}{
	\includegraphics[width=\linewidth]{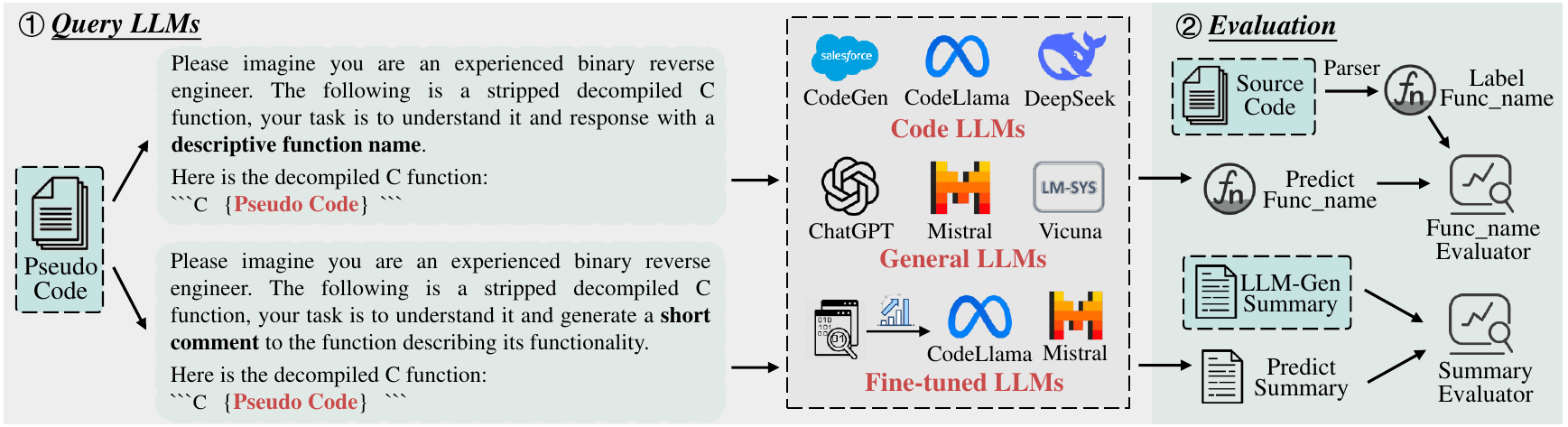}
         }
	\caption{An overview of the evaluation process.}
    \label{fig:eva_overview}
\end{figure*}

\subsubsection{\textbf{Prompt Formats}}
\vspace{-1ex}
Figure \ref{fig:eva_overview} illustrates the prompt format we used for LLMs. We use role-play \citep{chen2023unleashing, kong2024better} prompts to give LLMs the role of experienced binary reverse engineers, enabling them to better handle binary code understanding tasks.
We enclose the code in the prompt with triple backticks (\textasciigrave\textasciigrave\textasciigrave) to clearly describe the code format.
Considering the limitation of the length of the model context window, and in order to reduce the inference time overhead and memory usage, we adopt the zero-shot prompts. We also analyze the impact of few-shot prompts on the performance of LLMs in Section \ref{sec:rq3}.

\subsubsection{\textbf{Fine-tuned Large Language Models}} \label{subsec:finetunellm}
\vspace{-1ex}
We also investigate the ability of fine-tuned LLMs to understand binary code, since fine-tuning is a common technique to adapt a pre-trained LLM to downstream tasks \citep{HowEffective_ISSTA_2023,feng2020codebert,fried2023incoder}, such as function name recovery and code summary generation. Furthermore, pre-training corpora of existing LLMs contain very few binary code, either in the form of disassembled instruction sequences or decompiled pseudo code. Therefore, we hope to explore whether injecting binary domain information into LLMs can improve its performance.

The GNU repository\footnote{\url{http://ftp.gnu.org/gnu}} is extensively used as a training or test set for many existing deep learning-based works \citep{nero2020, symlm2022ccs, Chen2023pst, Xiong2023ase,jTrans2022wang, PalmTree21li}. To build our fine-tuning dataset, we select 51 projects from the GNU repository, including binutils, coreutils, findutils, libmicrohttpd, nettle, etc. We use BinKit \citep{Kim2023tse} to create a compilation environment, and then compiler the selected projects using the \texttt{gcc-11.2.0} compiler with \texttt{x86\_64} target architecture and \texttt{O0} optimization level. We obtain a total of 270 binary files. After strip, decompile and alignment, we obtain 124,819 functions matching source code and decompiled pseudo code, and randomly select 30,000 of them as the fine-tuning dataset.

Additionally, we perform a search and confirm that none of the functions in our proposed benchmark is present in the fine-tuning dataset.

\subsection{\textbf{Evaluation Setup}}
The evaluation environment is a machine equipped with 8 * NVIDIA RTX A6000 GPU with 48GB of VRAM, 2 * 28-core Intel Xeon 6330 CPU, 512GB RAM and 64TB storage, running on Ubuntu 22.04 OS. The GPU is running Nvidia driver version 525.116.03 with CUDA version 12.0. 

We implement all the experiments using Python 3.8 with PyTorch \citep{PyTorch} 2.0.1, DeepSpeed \citep{DeepSpeed} 0.13.0 and Transformers \citep{Transformers} 4.37.2 packages. As for model fine-tuning, we implement it based on the LLaMa-Factory \citep{zheng2024llamafactory} framework. We use LoRA \citep{hu2021lora} fine-tuning method and specify all available modules. We adopt Adam optimizer in fp16 precision, 40 global batch size and 1 training epoch. The learning rate is set to 1e-5 and followed by a consine decay.

\section{Evaluation Results and Findings \label{sec:results}}

In this section, we conduct extensive experiments to answer the following research questions:

\begin{itemize}
\item \textbf{RQ1:} How do LLMs perform in the task of function name recovery?
\item \textbf{RQ2:} How do LLMs perform in the task of binary code summarization?
\item \textbf{RQ3:} What factors significantly influence the performance of LLMs to understand binary code?
\item \textbf{RQ4:} Can fine-tuning enhance the capability of LLMs to understand binary code?
\item \textbf{RQ5:} Do LLMs have the practical ability in real-world scenarios?
\end{itemize}

\subsection{\textbf{RQ1: Performance of Function Name Recovery \label{sec:rq1}}}

\begin{sidewaystable*}[thp]
  \vspace{100ex}
  \caption{Comparison of LLMs and DL-based methods on function name recovery under different target architectures. We mark the \colorbox{mycolor}{best} performing methods in each domain.}
  \centering
  \renewcommand{\arraystretch}{1.20}
  \setlength{\tabcolsep}{0.6mm}{
  \scalebox{0.90}{
    \begin{tabular}{c@{\hspace{0pt}}l@{\hspace{0pt}}cccccccccccccccc}
    \toprule
     \multirow{2}{*}{\textbf{Domain}} & \multicolumn{1}{c}{\multirow{2}{*}{\textbf{Model}}} 
     & \multicolumn{4}{c}{\textbf{x64}} & \multicolumn{4}{c}{\textbf{x86}} & \multicolumn{4}{c}{\textbf{ARM}} & \multicolumn{4}{c}{\textbf{MIPS}}   \\
     \cmidrule(r){3-6} \cmidrule(r){7-10}  \cmidrule(r){11-14} \cmidrule(r){15-18}
     & & \textbf{\emph{Precision}} & \textbf{\emph{Recall}} & \textbf{\emph{F1-score}} & \textbf{\emph{Time(s)}} & \textbf{\emph{Precision}} & \textbf{\emph{Recall}} & \textbf{\emph{F1-score}} & \textbf{\emph{Time(s)}} & \textbf{\emph{Precision}} & \textbf{\emph{Recall}} & \textbf{\emph{F1-score}} & \textbf{\emph{Time(s)}} & \textbf{\emph{Precision}} & \textbf{\emph{Recall}} & \textbf{\emph{F1-score}} & \textbf{\emph{Time(s)}}   \\
    \midrule
     \multicolumn{1}{c}{\multirow{8}{*}{\makecell{Code \\ LLMs}}} 
       & CodeGen25-7b-instruct & 9.98 & 12.62 & 10.06 & 1.63 & 9.68 & 12.46 & 9.89 & 1.47 & 10.07 & 12.77 & 10.13 & 1.47 & 14.01 & 16.25 & 13.89 & 1.48 \\
       & WizardCoder-15b-V1.0 & 23.29 & 22.11 & 21.76 & 1.57 & 23.66 & 23.03 & 22.45 & 1.55 & 23.93 & 23.32 & 22.64 & 1.53 & 29.58 & 28.80 & 28.09 & 1.51 \\
       & WizardCoder-33b-V1.1 & 20.57 & 21.13 & 19.49 & 5.01 & 20.79 & 20.85 & 19.70 & 4.79 & 20.54 & 20.09 & 19.18 & 4.85 & 25.04 & 24.71 & 23.47 & 4.81  \\
       & Code Llama-7b-instruct-hf & 27.52 & 25.06 & 25.23 & \cellcolor{mycolor}{1.19} & 26.73 & 24.76 & 24.81 & \cellcolor{mycolor}{1.40} & 26.14 & 25.18 & 24.58 & \cellcolor{mycolor}{1.38} & 31.27 & 29.77 & 29.36 & \cellcolor{mycolor}{1.35} \\
       & Code Llama-13b-instruct-hf & 26.64 & 24.83 & 24.68 & 1.72 & 25.44 & 23.84 & 23.63 & 1.68 & 26.41 & 24.84 & 24.67 & 1.69 & 31.35 & 30.04 & 29.57 & 1.68  \\
       & Code Llama-34b-instruct-hf & \cellcolor{mycolor}{28.73} & \cellcolor{mycolor}{27.33} & \cellcolor{mycolor}{26.75} & 3.42 & \cellcolor{mycolor}{28.06} & \cellcolor{mycolor}{26.82} & \cellcolor{mycolor}{26.09} & 3.43 & \cellcolor{mycolor}{29.14} & \cellcolor{mycolor}{27.83} & \cellcolor{mycolor}{27.16} & 3.33 & \cellcolor{mycolor}{32.87} & \cellcolor{mycolor}{31.22} & \cellcolor{mycolor}{30.66} & 3.40 \\
       & Code Llama-70b-instruct-hf & 26.57 & 25.52 & 24.90 & 10.05 & 26.88 & 25.75 & 25.00 & 10.93 & 26.79 & 25.78 & 25.16 & 9.83 & 30.03 & 29.61 & 28.55 & 9.94 \\
       & DeepSeek-Coder-33b-instruct & 23.56 & 24.58 & 23.02 & 3.99 & 22.59 & 22.99 & 21.83 & 4.01 & 23.74 & 24.09 & 22.84 & 3.83 & 27.80 & 28.68 & 27.05 & 3.91 \\
    \midrule
      \multicolumn{1}{c}{\multirow{8}{*}{\makecell{General \\ LLMs}}} 
       & ChatGLM2-6B & 13.88 & 12.57 & 12.66 & 1.08 & 15.26 & 13.48 & 13.74 & 1.10 & 13.96 & 12.80 & 12.81 & \cellcolor{mycolor}{1.02} & 17.63 & 16.18 & 16.15 & \cellcolor{mycolor}{1.01} \\
       & Vicuna-7b-v1.5 & 18.28 & 18.17 & 17.32 & 1.13 & 17.81 & 16.97 & 16.68 & 1.06 & 17.71 & 17.57 & 16.86 & 1.06 & 23.03 & 23.09 & 22.07 & \cellcolor{mycolor}{1.01} \\
       & Vicuna-13b-v1.5 & 21.24 & 21.54 & 20.46 & \cellcolor{mycolor}{1.03} & 20.56 & 21.13 & 19.96 & \cellcolor{mycolor}{1.03} & 20.94 & 21.81 & 20.37 & 1.04 & 24.22 & 25.19 & 23.57 & 1.05 \\
       & Llama-2-13b-chat-hf & \cellcolor{mycolor}{23.88} & 23.28 & \cellcolor{mycolor}{22.68} & 1.18 & \cellcolor{mycolor}{23.90} & 23.03 & \cellcolor{mycolor}{22.54} & 1.39 & \cellcolor{mycolor}{23.98} & 23.03 & \cellcolor{mycolor}{22.51} & 1.25 & \cellcolor{mycolor}{28.17} & 27.77 & \cellcolor{mycolor}{26.88} & 1.17 \\
       & Llama-2-70b-chat-hf & 23.51 & 21.56 & 21.72 & 4.46 & 22.88 & 21.23 & 21.12 & 4.26 & 22.68 & 21.29 & 21.16 & 4.27 & 26.14 & 24.61 & 24.43 & 4.35  \\
       & Mistral-7B-Instruct-v0.2 & 21.92 & 23.82 & 21.81 & 1.20 & 20.87 & 22.61 & 20.72 & 1.26 & 21.62 & 23.22 & 21.34 & 1.20 & 25.57 & 27.43 & 25.22 & 1.32 \\
       & Mixtral-8x7B-Instruct-v0.1 & 19.34 & \cellcolor{mycolor}{25.65} & 21.00 & 2.65 & 18.97 & \cellcolor{mycolor}{25.55} & 20.72 & 2.65 & 19.58 & \cellcolor{mycolor}{26.12} & 21.42 & 2.73 & 21.50 & \cellcolor{mycolor}{29.10} & 23.58 & 2.69 \\
       & ChatGPT(gpt-3.5-turbo-16k) & 19.40 & 18.41 & 18.08 & - & 18.23 & 20.81 & 18.66 & - & 18.66 & 21.10 & 19.03 & - & 20.90 & 23.71 & 21.39 & - \\
    \midrule
    \multicolumn{1}{c}{\multirow{3}{*}{DL-based}} 
        & SymLM \citep{symlm2022ccs} & 9.63 & 5.54 & 6.66 & 0.28  & 9.55 & 5.39 & 6.59 & 0.29 & 9.61 & 5.50 & 6.68 & 0.28 & 10.14 & 5.76 & 6.78 & 0.27\\
        & NER \citep{Chen2023pst} & \cellcolor{mycolor}{16.22} & \cellcolor{mycolor}{9.83} & \cellcolor{mycolor}{11.17} & \cellcolor{mycolor}{0.03} & \cellcolor{mycolor}{16.31} & \cellcolor{mycolor}{10.03} & \cellcolor{mycolor}{11.40} & \cellcolor{mycolor}{0.03} & \cellcolor{mycolor}{15.99} & \cellcolor{mycolor}{9.49} & \cellcolor{mycolor}{10.87} & \cellcolor{mycolor}{0.03} & \cellcolor{mycolor}{16.98} & \cellcolor{mycolor}{10.36} & \cellcolor{mycolor}{11.73} & \cellcolor{mycolor}{0.03} \\
        & HexT5 \citep{Xiong2023ase} & 3.39 & 2.97 & 3.00 & 0.17 & 3.44 & 3.10 & 3.10 & 0.16 & 3.33 & 3.01 & 2.95 & 0.16 & 3.59 & 3.24 & 3.31 & 0.18\\
    \bottomrule
    \end{tabular} } }
    \vspace{0ex}
  \label{tab:funcname_result_architecture}%
\end{sidewaystable*}

\noindent\textbf{Metrics.} For the function name recovery task, following \citep{nfre2021issta, Chen2023pst}, we calculate token-level Precision, Recall, F1-score to evaluate the performance of LLMs. The metrics ignore non-alphabetical characters and are case-, order-, and duplication-insensitive at the token-level. They can be expressed as:
\begin{equation}
\begin{aligned}
\text{Precision}=\frac{\text{TP}}{\text{TP + FP}}
\end{aligned}
\end{equation}
\begin{equation}
\begin{aligned}
\text{Recall}=\frac{\text{TP}}{\text{TP + FN}}
\end{aligned}
\end{equation}
\begin{equation}
\begin{aligned}
\text{F1-score}=\frac{2 \times \text{Precision} \times \text{Recall}}{\text {Precision}+\text{Recall}}
\end{aligned}
\end{equation}
Specifically, function names consist of one or more discrete tokens, each encapsulating a portion of the function’s semantic information. During the evaluation, we employ a combination of empirical rules and a unigram language model (ULM)\citep{ulm} to segment function names into token sequences. For function names that adhere to standard naming conventions, such as camel case (\texttt{getUserId}) and snake case (\texttt{get\_user\_id}), we design specific rules for tokenization. However, for function names that do not follow explicit naming conventions, such as \texttt{getuserid}, we utilize a pretrained SentencePiece model in conjunction with ULM to segment names based on token frequency statistics. For instance, \texttt{setcmdfmt} is tokenized into \texttt{set}, \texttt{cmd}, and \texttt{fmt}.

For an intuitive understanding of the metrics, a ground truth of \texttt{getUserMessage} with a prediction of \texttt{get\_message} is given full precision and 67\% recall. And a prediction of \texttt{get\_user\_message\_set\_user\_message} has 83\% precision and full recall.

\noindent\textbf{Results.} Table~\ref{tab:funcname_result_architecture}  and Table~\ref{tab:funcname_result_optimization} respectively show the performance of LLMs and DL-based methods in function name recovery tasks under different target architectures and compiler optimization options.

From Table~\ref{tab:funcname_result_architecture}, we can observe that CodeLlama and Llama consistently perform excellently across different parameter sizes under various architectures. Among them, CodeLlama-34b outperforms all other LLMs in precision, recall, and F1-score metrics, achieving scores of 28.73, 27.33, and 26.75 in the \texttt{x64} architecture, 28.06, 26.82, and 26.09 in the \texttt{x86} architecture, 29.14, 27.83, and 27.16 in the \texttt{ARM} architecture, and 32.87, 31.22, and 30.66 in the \texttt{MIPS} architecture. Following closely behind, WizardCoder-15b and DeepSeek-Coder-33b also show strong performance, with F1-score of 21.76, 22.45, 22.64, 28.09 and 23.02, 21.83, 22.84, 27.05 across the four architectures, respectively. CodeGen25-7b and ChatGLM2-6B show the poorest performance, with F1-score ranging from only 48.9\% to 66.7\% of the other LLMs. This lagging performance may be attributed to the capability flaws of their basic models or the lack of targeted training data. In the general LLMs category, Llama-2-13b consistently outperforms other LLMs in both precision and F1-score metrics, achieving scores of 23.88 and 22.68 in the \texttt{x64} architecture, 23.90 and 22.54 in the \texttt{x86} architecture, 23.98 and 22.51 in the \texttt{ARM} architecture, and 28.17 and 26.88 in the \texttt{MIPS} architecture, respectively.

A deeper analysis of the performance across different target architectures reveals that the average F1-score for all LLMs on the \texttt{MIPS} architecture is 24.62, while on \texttt{x64}, \texttt{x86}, \texttt{ARM} architectures, the average F1-score are 20.73, 20.47, and 20.74, respectively. Compared to other architectures, the F1-score for \texttt{MIPS} shows improvements of 18.77\%, 20.27\%, and 18.71\%. The superior performance of \texttt{MIPS} in function name recovery tasks can likely be attributed to its simplified instruction set and unified function calling conventions. This streamlined instruction set reduces reliance on complex operations and diverse instruction variants, enabling the LLM to more effectively capture function call patterns and contextual information.

Furthermore, as a task that combines natural language and code language understanding, code domain LLMs generally perform slightly better than general domain LLMs. This may be because their pre-training datasets have a higher proportion of source code. Training on a extensive range of source code datasets allows them to gain a deeper grasp of programming syntax, structure and semantics.

\finding{CodeLlama-34b performs the best in the function name recovery task across different target architectures, achieving an F1-score of 26.75, 26.35, 26.75, and 26.59 in the \texttt{x64},  \texttt{x86},  \texttt{ARM}, and \texttt{MIPS} architectures, respectively. Additionally, LLMs perform the best in function name recovery on the \texttt{MIPS} architecture, with the average F1-score across all LLMs being 24.62. Code domain LLMs generally perform slightly better than general domain LLMs, likely owing to their greater familiarity with programming paradigms.}

\begin{sidewaystable*}[thp]
  \vspace{100ex}
  \caption{Comparison of LLMs and DL-based methods on function name recovery under different compiler optimization options. We mark the \colorbox{mycolor}{best} performing methods in each domain.}
  \centering
  \renewcommand{\arraystretch}{1.20}
  \setlength{\tabcolsep}{0.6mm}{
  \scalebox{0.90}{
    \begin{tabular}{c@{\hspace{0pt}}l@{\hspace{0pt}}cccccccccccccccc}
    \toprule
     \multirow{2}{*}{\textbf{Domain}} & \multicolumn{1}{c}{\multirow{2}{*}{\textbf{Model}}} 
     & \multicolumn{4}{c}{\textbf{O0}} & \multicolumn{4}{c}{\textbf{O1}} & \multicolumn{4}{c}{\textbf{O2}} & \multicolumn{4}{c}{\textbf{O3}}   \\
     \cmidrule(r){3-6} \cmidrule(r){7-10}  \cmidrule(r){11-14} \cmidrule(r){15-18}
     & & \textbf{\emph{Precision}} & \textbf{\emph{Recall}} & \textbf{\emph{F1-score}} & \textbf{\emph{Time(s)}} & \textbf{\emph{Precision}} & \textbf{\emph{Recall}} & \textbf{\emph{F1-score}} & \textbf{\emph{Time(s)}} & \textbf{\emph{Precision}} & \textbf{\emph{Recall}} & \textbf{\emph{F1-score}} & \textbf{\emph{Time(s)}} & \textbf{\emph{Precision}} & \textbf{\emph{Recall}} & \textbf{\emph{F1-score}} & \textbf{\emph{Time(s)}}   \\
    \midrule
     \multicolumn{1}{c}{\multirow{8}{*}{\makecell{Code \\ LLMs}}} 
       & CodeGen25-7b-instruct & 9.98 & 12.62 & 10.06 & 1.63 & 9.83 & 12.25 & 9.85 & 1.44 & 10.02 & 12.78 & 10.36 & 1.54 & 10.23 & 12.89 & 10.77 & \cellcolor{mycolor}{1.56} \\
       & WizardCoder-15b-V1.0 & 23.29 & 22.11 & 21.76 & 1.57 & 24.24 & 23.45 & 22.86 & 1.67 & 24.44 & 23.82 & 23.13 & 1.58 & 24.58 & 23.82 & 23.16 & 1.62   \\
       & WizardCoder-33b-V1.1 & 20.57 & 21.13 & 19.49 & 5.01 & 20.20 & 20.90 & 19.23 & 5.09 & 20.60 & 21.52 & 19.63 & 5.22 & 20.76 & 21.18 & 19.74 & 5.30 \\
       & Code Llama-7b-instruct-hf & 27.52 & 25.06 & 25.23 & \cellcolor{mycolor}{1.19} & 26.28 & 24.75 & 24.55 & \cellcolor{mycolor}{1.28} & 25.43 & 24.06 & 23.81 & \cellcolor{mycolor}{1.25} & 26.64 & 24.82 & 24.74 & 1.57 \\
       & Code Llama-13b-instruct-hf & 26.64 & 24.83 & 24.68 & 1.72 & 26.59 & 24.47 & 24.45 & 1.92 & 26.56 & 24.50 & 24.53 & 1.81 & 27.33 & 24.99 & 25.10 & 1.88  \\
       & Code Llama-34b-instruct-hf & \cellcolor{mycolor}{28.73} & \cellcolor{mycolor}{27.33} & \cellcolor{mycolor}{26.75} & 3.42 & \cellcolor{mycolor}{27.98} & \cellcolor{mycolor}{27.50} & \cellcolor{mycolor}{26.35} & 3.59 & \cellcolor{mycolor}{28.89} & \cellcolor{mycolor}{27.09} & \cellcolor{mycolor}{26.75} & 3.57 & \cellcolor{mycolor}{28.40} & \cellcolor{mycolor}{27.31} & \cellcolor{mycolor}{26.59} & 3.77 \\
       & Code Llama-70b-instruct-hf & 26.57 & 25.52 & 24.90 & 10.05 & 26.45 & 25.93 & 25.10 & 9.97 & 26.59 & 25.57 & 24.90 & 10.10 & 26.41 & 25.47 & 24.80 & 10.35 \\
       & DeepSeek-Coder-33b-instruct & 23.56 & 24.58 & 23.02 & 3.99 & 23.69 & 24.92 & 23.21 & 4.11 & 23.42 & 24.39 & 22.87 & 4.21 & 23.76 & 24.95 & 23.25 & 4.22 \\
    \midrule
      \multicolumn{1}{c}{\multirow{8}{*}{\makecell{General \\ LLMs}}} 
       & ChatGLM2-6B & 13.88 & 12.57 & 12.66 & 1.08 & 13.75 & 12.20 & 12.39 & \cellcolor{mycolor}{1.04} & 13.34 & 12.00 & 12.11 & \cellcolor{mycolor}{1.02} & 14.10 & 12.42 & 12.67 & \cellcolor{mycolor}{1.04} \\
       & Vicuna-7b-v1.5 & 18.28 & 18.17 & 17.32 & 1.13 & 18.36 & 17.37 & 17.03 & 1.10 & 17.50 & 17.27 & 16.63 & 1.15 & 18.72 & 17.80 & 17.44 & 1.13 \\
       & Vicuna-13b-v1.5 & 21.24 & 21.54 & 20.46 & \cellcolor{mycolor}{1.03} & 20.29 & 20.71 & 19.62 & 1.10 & 20.50 & 20.80 & 19.70 & 1.09 & 20.11 & 20.55 & 19.33 & 1.18 \\
       & Llama-2-13b-chat-hf & \cellcolor{mycolor}{23.88} & 23.28 & \cellcolor{mycolor}{22.68} & 1.18 & \cellcolor{mycolor}{24.06} & 23.20 & \cellcolor{mycolor}{22.71} & 1.18 & \cellcolor{mycolor}{24.44} & 23.48 & \cellcolor{mycolor}{23.00} & 1.20 & \cellcolor{mycolor}{24.42} & 23.62 & \cellcolor{mycolor}{23.04} & 1.30 \\
       & Llama-2-70b-chat-hf & 23.51 & 21.56 & 21.72 & 4.46 & 22.68 & 20.95 & 20.99 & 4.45 & 22.15 & 20.50 & 20.53 & 4.52 & 22.93 & 20.87 & 21.02 & 4.81 \\
       & Mistral-7B-Instruct-v0.2 & 21.92 & 23.82 & 21.81 & 1.20 & 21.53 & 23.97 & 21.62 & 1.26 & 21.38 & 23.47 & 21.37 & 1.33 & 21.44 & 23.37 & 21.36 & 1.36  \\
       & Mixtral-8x7B-Instruct-v0.1 & 19.34 & \cellcolor{mycolor}{25.65} & 21.00 & 2.65 & 19.09 & \cellcolor{mycolor}{25.76} & 20.82 & 2.69 & 18.73 & \cellcolor{mycolor}{25.33} & 20.57 & 2.75 & 19.52 & \cellcolor{mycolor}{25.86} & 21.15 & 2.79 \\
       & ChatGPT(gpt-3.5-turbo-16k) &  19.40 & 18.41 & 18.08 & - & 18.63 & 21.58 & 19.21 & - & 18.12 & 21.00 & 18.65 & - & 18.36 & 21.12 & 18.85 & - \\
    \midrule
    \multicolumn{1}{c}{\multirow{3}{*}{DL-based}} 
        & SymLM \citep{symlm2022ccs} & 9.63 & 5.54 & 6.66 & 0.28 & 9.58 & 5.39 & 6.48 & 0.26 & 9.51 & 5.42 & 6.51 & 0.26 & 9.48 & 5.30 & 6.44 & 0.25 \\
        & NER \citep{Chen2023pst} & \cellcolor{mycolor}{16.22} & \cellcolor{mycolor}{9.83} & \cellcolor{mycolor}{11.17} & \cellcolor{mycolor}{0.03}  & \cellcolor{mycolor}{16.10} & \cellcolor{mycolor}{9.41} & \cellcolor{mycolor}{10.99} & \cellcolor{mycolor}{0.03}  & \cellcolor{mycolor}{16.17} & \cellcolor{mycolor}{9.62} & \cellcolor{mycolor}{11.01} & \cellcolor{mycolor}{0.03}  & \cellcolor{mycolor}{16.28} & \cellcolor{mycolor}{9.91} & \cellcolor{mycolor}{11.33} & \cellcolor{mycolor}{0.03} \\
        & HexT5 \citep{Xiong2023ase} & 3.39 & 2.97 & 3.00 & 0.17 & 3.30 & 2.88 & 2.85 & 0.18 & 3.24 & 2.82 & 2.91 & 0.17 & 3.38 & 3.01 & 3.09 & 0.16\\
    \bottomrule
    \end{tabular} } }
    \vspace{0ex}
  \label{tab:funcname_result_optimization}%
\end{sidewaystable*}

From Table~\ref{tab:funcname_result_optimization}, it is evident that CodeLlama and Llama also continue to perform excellently under different compiler optimization options. Among them, CodeLlama-34b outperforms all other LLMs in precision, recall, and F1-score metrics, achieving scores of 28.73, 27.33, and 26.75 under the \texttt{-O0} optimization, 27.98, 27.50, and 26.35 under the \texttt{-O1} optimization, 28.89, 27.09, and 26.75 under the \texttt{-O2} optimization, and 28.40, 27.31, and 26.59 under the \texttt{-O3} optimization. 

In the general LLM category, Llama-2-13b outperforms other general LLMs in both precision and F1-score, achieving the following scores across the four optimization options: 23.88 and 22.68 for \texttt{-O0}, 24.06 and 22.71 for \texttt{-O1}, 24.44 and 23.00 for \texttt{-O2}, and 24.42 and 23.04 for \texttt{-O3}, respectively.

For the binaries with \texttt{-O0}, \texttt{-O1}, \texttt{-O2}, and \texttt{-O3} optimization levels, the average F1-scores for all LLMs are 20.73, 20.62, 20.53, and 20.81, respectively. We observe that the performance differences in function name recovery tasks across different optimization levels are minimal, with the performance gap between \texttt{O0} and \texttt{O3} being only 0.39\%.

\finding{CodeLlama-34b continues to perform the best in the function name recovery task under different compiler optimization options, achieving an F1-score of 26.75, 26.35, 26.75, and 26.59 under the \texttt{-O0}, \texttt{-O1}, \texttt{-O2}, and \texttt{-O3} optimization levels, respectively. Additionally, all LLMs exhibit a minimal performance gap across different optimization levels, with the average F1-score difference between \texttt{O0} and \texttt{O3} being only 0.39\%.}

Among the deep learning-based expert models, NER \citep{Chen2023pst} performs the best, with an average F1-score of 11.29 and 11.13 in different architectures and optimization options, still slightly outperforming the LLM with the worst performance, CodeGen25. However, SymLM \citep{symlm2022ccs} and HexT5 \citep{Xiong2023ase} perform poorly, with average F1-score of 3.68 and 6.52, and 3.09 and 2.96, respectively, which differ from the performance reported in their original papers. This difference may come from the partitioning of their datasets. SymLM and HexT5 widely use projects from GNU as part of their training and testing sets. SymLM divides the training and testing sets at the binary file-level, which may result in some code appearing in both the training and testing sets. For example, in the \texttt{Binutils} project, the \texttt{ar} and \texttt{nm} files share the same binary file descriptor (BFD) processing code. This reuse of libraries and underlying code may lead to exaggeration of evaluation metrics. Although HexT5 adopts a stricter project-level dataset partitioning approach, different projects under GNU may still share programming styles or naming conventions, leading to potential data leaks. Overall, deep learning-based expert models perform worse compared to LLMs, primarily due to their limited generalization ability, which makes them prone to overfitting specific features of the training data when handling out-of-distribution data. In contrast, LLMs exhibit superior adaptability in zero-shot learning tasks, allowing them to effectively handle unseen data.

\finding{Among the existing deep learning-based expert models, SymLM performs the best. However, these models exhibit poor generalization ability beyond the training data distribution, with performance significantly lower than that of LLMs.}

In terms of inference time cost, locally deployed LLMs with 6B-7B parameter quantities typically require 1 to 1.6 seconds to infer a single piece of data, LLMs with 13-15B scales require 1.2 to 2 seconds, LLMs with 33-34B scales require 3.5 to 5.1 seconds, and CodeLlama-70b requires a maximum of 10.35 seconds per piece. 
Considering that ChatGPT requires API access, which is affected by network latency and rate limits, we do not measure its time overhead.
The DL-based model, due to its lightweight advantage, greatly reduces inference time and achieves the fastest NER of 0.03 seconds per piece.

\finding{The DL-based model has a significant advantage in inference speed benefiting from their model size. Meanwhile, the inference speed of LLMs is still within an acceptable range.}

\begin{sidewaystable*}[thp]
  \vspace{100ex}
  \caption{Comparison of LLMs and DL-based methods on binary code summarization under different target architectures. We mark the \colorbox{mycolor}{best} performing methods in each domain.}
  \centering
  \renewcommand{\arraystretch}{1.20}
  \setlength{\tabcolsep}{0.5mm}{
  \scalebox{0.86}{
    \begin{tabular}{c@{\hspace{0pt}}l@{\hspace{0pt}}cccccccccccccccc}
    \toprule
     \multirow{2}{*}{\textbf{Domain}} & \multicolumn{1}{c}{\multirow{2}{*}{\textbf{Model}}} 
     & \multicolumn{4}{c}{\textbf{x64}} & \multicolumn{4}{c}{\textbf{x86}} & \multicolumn{4}{c}{\textbf{ARM}} & \multicolumn{4}{c}{\textbf{MIPS}}   \\
     \cmidrule(r){3-6} \cmidrule(r){7-10}  \cmidrule(r){11-14} \cmidrule(r){15-18}
     & & \textbf{\emph{BLEU-4}} & \textbf{\emph{METEOR}} & \textbf{\emph{Rouge-L}} & \textbf{\emph{Time(s)}} & \textbf{\emph{BLEU-4}} & \textbf{\emph{METEOR}} & \textbf{\emph{Rouge-L}} & \textbf{\emph{Time(s)}} & \textbf{\emph{BLEU-4}} & \textbf{\emph{METEOR}} & \textbf{\emph{Rouge-L}} & \textbf{\emph{Time(s)}} & \textbf{\emph{BLEU-4}} & \textbf{\emph{METEOR}} & \textbf{\emph{Rouge-L}} & \textbf{\emph{Time(s)}} \\
    \midrule
     \multicolumn{1}{c}{\multirow{8}{*}{\makecell{Code \\ LLMs}}} 
       & CodeGen25-7b-instruct  & 3.56 & 20.76 & 14.06 & 7.14 & 3.55 & 20.99 & 14.11 & 7.19 & 3.53 & 20.97 & 14.38 & 7.17 & 3.79 & 21.71 & 14.61 & 7.15 \\
       & WizardCoder-15b-V1.0  & \cellcolor{mycolor}{4.72} & \cellcolor{mycolor}{24.10} & \cellcolor{mycolor}{18.16} & 8.00 & \cellcolor{mycolor}{4.63} & \cellcolor{mycolor}{24.04} & \cellcolor{mycolor}{18.12} & 7.95 & \cellcolor{mycolor}{4.67} & \cellcolor{mycolor}{24.23} & \cellcolor{mycolor}{18.02} & 8.16 & \cellcolor{mycolor}{5.08} & \cellcolor{mycolor}{25.16} & \cellcolor{mycolor}{19.07} & 7.85 \\
       & WizardCoder-33b-V1.1  & 4.12 & 23.72 & 16.03 & 20.77 & 4.07 & 23.70 & 15.84 & 21.25 & 4.05 & 23.68 & 16.01 & 20.92 & 4.38 & 24.72 & 16.63 & 20.88 \\
       & Code Llama-7b-instruct-hf  & 4.16 & 21.10 & 16.65 & \cellcolor{mycolor}{5.55} & 4.06 & 21.18 & 16.62 & \cellcolor{mycolor}{5.42} & 4.15 & 20.92 & 16.74 & \cellcolor{mycolor}{5.16} & 4.47 & 21.80 & 17.49 & \cellcolor{mycolor}{5.42} \\
       & Code Llama-13b-instruct-hf  & 4.22 & 19.52 & 16.46 & 7.62 & 4.11 & 19.65 & 16.24 & 7.15 & 4.14 & 19.74 & 16.54 & 7.02 & 4.35 & 20.22 & 16.91 & 6.98 \\
       & Code Llama-34b-instruct-hf & 4.41 & 20.97 & 17.60 & 15.70 & 4.45 & 20.96 & 17.61 & 16.37 & 4.34 & 20.98 & 17.64 & 16.29 & 4.77 & 21.44 & 18.41 & 15.44 \\
       & Code Llama-70b-instruct-hf & 4.26 & 22.37 & 16.60 & 46.33 & 4.07 & 22.26 & 16.37 & 46.73 & 4.19 & 22.36 & 16.63 & 46.38 & 4.45 & 23.13 & 17.30 & 47.21 \\
       & DeepSeek-Coder-33b-instruct  & 4.66 & 24.08 & 17.38 & 17.32 & 4.53 & 23.97 & 17.13 & 17.98 & 4.64 & 24.22 & 17.37 & 17.36 & 5.05 & 25.11 & 18.18 & 17.15 \\
    \midrule
      \multicolumn{1}{c}{\multirow{8}{*}{\makecell{General \\ LLMs}}} 
       & ChatGLM2-6B & 3.54 & 21.42 & 14.49 & 5.21 & 3.55 & 21.62 & 14.69 & 5.14 & 3.54 & 21.78 & 14.72 & 5.05 & 3.80 & 22.40 & 15.19 & 5.00 \\
       & Vicuna-7b-v1.5  & 4.47 & 23.02 & 17.64 & 4.40 & 4.29 & 22.94 & 17.34 & 4.49 & 4.39 & 23.26 & 17.79 & 4.39 & 4.79 & 24.03 & 18.50 & 4.06 \\
       & Vicuna-13b-v1.5 & 5.59 & 22.23 & 19.75 & 4.86 & 5.47 & 22.34 & 19.78 & 5.20 & 5.55 & 22.38 & 19.87 & 5.20 & 6.06 & 23.36 & 20.86 & 5.10 \\
       & Llama-2-13b-chat-hf  & 4.72 & 22.56 & 18.06 & 7.07 & 4.79 & 22.67 & 18.35 & 7.21 & 4.86 & 22.92 & 18.61 & 6.85 & 5.28 & 24.02 & 19.58 & 6.72 \\
       & Llama-2-70b-chat-hf & 4.36 & 23.17 & 16.14 & 45.73 & 4.35 & 23.50 & 16.26 & 44.49 & 4.50 & 23.93 & 16.66 & 42.85 & 4.82 & 24.76 & 17.39 & 42.60 \\
       & Mistral-7B-Instruct-v0.2  & 6.05 & 25.65 & 20.86 & \cellcolor{mycolor}{3.51} & 5.74 & 25.53 & 20.62 & \cellcolor{mycolor}{3.53} & 6.11 & 25.33 & 21.06 & \cellcolor{mycolor}{3.55} & 6.62 & 26.40 & 22.08 & \cellcolor{mycolor}{3.43} \\
       & Mixtral-8x7B-Instruct-v0.1 & 6.41 & 26.02 & 21.04 & 12.41 & 6.16 & 25.34 & 20.69 & 11.23 & 6.47 & 25.86 & 21.19 & 13.63 & 6.77 & 26.88 & 21.97 & 10.71 \\
       & ChatGPT(gpt-3.5-turbo-16k) & \cellcolor{mycolor}{7.69} & \cellcolor{mycolor}{29.50} & \cellcolor{mycolor}{22.09} & - & \cellcolor{mycolor}{7.38} & \cellcolor{mycolor}{28.65} & \cellcolor{mycolor}{21.52} & - & \cellcolor{mycolor}{7.80} & \cellcolor{mycolor}{29.48} & \cellcolor{mycolor}{22.48} & - & \cellcolor{mycolor}{8.30} & \cellcolor{mycolor}{30.32} & \cellcolor{mycolor}{23.13} & - \\
    \midrule
    \multicolumn{1}{c}{\multirow{2}{*}{DL-based}}
        & BinT5 \citep{bint52023saner}  & 0.00 & 2.08 & 4.69 & 0.56 & 0.00 & 2.00 & 4.61 & 0.57 & 0.00 & 2.03 & 4.67 & 0.58 & 0.00 & 2.11 & 4.82 & 0.57 \\
        & HexT5 \citep{Xiong2023ase}  & \cellcolor{mycolor}{0.10} & \cellcolor{mycolor}{6.21} & \cellcolor{mycolor}{8.44} & \cellcolor{mycolor}{0.43} & \cellcolor{mycolor}{0.09} & \cellcolor{mycolor}{6.13} & \cellcolor{mycolor}{8.37} & \cellcolor{mycolor}{0.44} & \cellcolor{mycolor}{0.09} & \cellcolor{mycolor}{6.09} & \cellcolor{mycolor}{8.33} & \cellcolor{mycolor}{0.44} & \cellcolor{mycolor}{0.11} & \cellcolor{mycolor}{6.29} & \cellcolor{mycolor}{8.53} & \cellcolor{mycolor}{0.42} \\
    \bottomrule
    \end{tabular} }
    \vspace{0ex}
  \label{tab:summarization_result_architecture} }
\end{sidewaystable*}

\subsection{\textbf{RQ2: Performance of Binary Code Summarization \label{sec:rq2}}}
\noindent\textbf{Metrics.} For the binary code summarization task, same as BinT5 \citep{bint52023saner}, HexT5 \citep{Xiong2023ase}, we use smoothed BLEU-4 \citep{papineni-etal-2002-bleu}, METEOR \citep{meteor2009Lavie}, Rouge-L \citep{lin-2004-rouge} as the evaluation metric.

BLEU (Bilingual Evaluation Understudy) is the most widely used metric in code summarization tasks. 
The Unigrams and bigrams measure the adequacy of the candidate, while longer trigrams and 4-grams assess its fluency. Based on standard works like CodeT5 \cite{wang-etal-2021-codet5} and CodeSearchNet \cite{husain2019codesearchnet}, we choose BLEU-4 as the evaluation metric.
BLEU-4 calculates the cumulative precision for 4-grams, which is the ratio of matching 4-grams in the candidate sentence to the total number of 4-grams. The score is computed as follows:
\begin{equation}
\small
\begin{aligned}
    BLEU\text{-}4 = BP \times \exp(\sum_{n=1}^{4}w_n \log P\_n),
\end{aligned}
\end{equation}
where $BP$ represents the brevity penalty for short generated sequences, $w_1$ to $w_n$ are positive weights summing to 1, and $P_n$ is the ratio of subsequences of length $n$ in the generated summary that also appear in the reference.

METEOR (Metric for Evaluation for Translation with Explicit Ordering) calculates the harmonic mean of the unigram precision and recall, which is calculated as:
\begin{equation}
\small
\begin{aligned}
    METEOR = (1-\gamma \cdot \textit{frag}^\beta) \cdot \frac{P \cdot R}{\alpha \cdot P + (1-\alpha) \cdot R},
\end{aligned}
\end{equation}
where \textit{frag} is the fragmentation fraction, and \emph{P} and \emph{R} represent unigram precision and recall, respectively. The parameters $\alpha$, $\beta$, and $\gamma$ are penalties.

Rouge-L is a variant of Rouge (Recall-oriented Understudy for Gisting Evaluation), which is calculated based on the longest common subsequence (LCS) betweem the reference and the candidate. The LCS-based F-measure ($F_{lcs}$) is called Rouge-L, which is calculated as:
\begin{equation}
\small
\begin{aligned}
    P_{lcs} = \frac{LCS(X,Y)}{n},\quad R_{lcs} = \frac{LCS(X,Y)}{m},\quad 
    F_{lcs} = \frac{(1+\beta ^2)P_{lcs}R_{lcs}}{R_{lcs}+\beta^2 P_{lcs}},
\end{aligned}
\end{equation}
where $\beta = P_{lcs}/R_{lcs}$, and \emph{n} and \emph{m} denote the lengths of \emph{X} and \emph{Y}, respectively.

\noindent\textbf{Results.} The performance of LLMs and deep learning-based methods in the binary code summarization task under different architectures and compiler optimization options is listed in Table~\ref{tab:summarization_result_architecture} and Table~\ref{tab:summarization_result_optimization}, respectively.

From Table~\ref{tab:summarization_result_architecture}, we can clearly observe that ChatGPT outperforms all other LLMs in BLEU-4, METEOR, and Rouge-L metrics across different architectures, achieving scores of 7.69, 29.50, and 22.09 in the \texttt{x64} architecture, 7.38, 28.65, and 21.52 in the \texttt{x86} architectures, 7.80, 29.48, and 22.48 in the \texttt{ARM} architectures, and 8.30, 30.32, and 23.13 in the \texttt{MIPS} architectures, respectively. At the same time, WizardCoder-15b also shows very competitive results, with scores of 4.72, 24.10, and 18.16 in the \texttt{x64} architecture, 4.63, 24.04, and 18.12 in the \texttt{x86} architecture, 4.67, 24.23, and 18.02 in the \texttt{ARM} architecture, and 5.08, 25.16, and 19.07 in the \texttt{MIPS} architecture. Similar to the function name recovery task, CodeGen25-7b and ChatGLM2-6B perform the worst in their respective domains, but narrow the performance gap with other LLMs.
Similar to the function name recovery task, all LLMs perform best on the \texttt{MIPS} architecture, with an average Rouge-L score of 18.58, compared to 17.69, 17.58, and 17.86 on the \texttt{x64},  \texttt{x86},  \texttt{ARM} architectures, respectively. The Rouge-L for the \texttt{MIPS} architecture is improved by 5.03\%, 5.69\%, and 4.03\% over the other architectures.

\begin{sidewaystable*}[thp]
  \vspace{100ex}
  \caption{Comparison of LLMs and DL-based methods on binary code summarization under different compiler optimization options. We mark the \colorbox{mycolor}{best} performing methods in each domain.}
  \centering
  \renewcommand{\arraystretch}{1.20}
  \setlength{\tabcolsep}{0.5mm}{
  \scalebox{0.86}{
    \begin{tabular}{c@{\hspace{0pt}}l@{\hspace{0pt}}cccccccccccccccc}
    \toprule
     \multirow{2}{*}{\textbf{Domain}} & \multicolumn{1}{c}{\multirow{2}{*}{\textbf{Model}}} 
     & \multicolumn{4}{c}{\textbf{O0}} & \multicolumn{4}{c}{\textbf{O1}} & \multicolumn{4}{c}{\textbf{O2}} & \multicolumn{4}{c}{\textbf{O3}}   \\
     \cmidrule(r){3-6} \cmidrule(r){7-10}  \cmidrule(r){11-14} \cmidrule(r){15-18}
     & & \textbf{\emph{BLEU-4}} & \textbf{\emph{METEOR}} & \textbf{\emph{Rouge-L}} & \textbf{\emph{Time(s)}} & \textbf{\emph{BLEU-4}} & \textbf{\emph{METEOR}} & \textbf{\emph{Rouge-L}} & \textbf{\emph{Time(s)}} & \textbf{\emph{BLEU-4}} & \textbf{\emph{METEOR}} & \textbf{\emph{Rouge-L}} & \textbf{\emph{Time(s)}} & \textbf{\emph{BLEU-4}} & \textbf{\emph{METEOR}} & \textbf{\emph{Rouge-L}} & \textbf{\emph{Time(s)}} \\
    \midrule
     \multicolumn{1}{c}{\multirow{8}{*}{\makecell{Code \\ LLMs}}} 
       & CodeGen25-7b-instruct  & 3.56 & 20.76 & 14.06 & 7.14 & 3.55 & 20.83 & 13.88 & 7.20 & 3.49 & 20.72 & 13.90 & 7.27 & 3.47 & 20.19 & 13.42 & 7.38 \\
       & WizardCoder-15b-V1.0  & \cellcolor{mycolor}{4.72} & \cellcolor{mycolor}{24.10} & \cellcolor{mycolor}{18.16} & 8.00 & \cellcolor{mycolor}{4.63} & 23.90 & \cellcolor{mycolor}{18.00} & 8.08 & \cellcolor{mycolor}{4.64} & \cellcolor{mycolor}{24.12} & \cellcolor{mycolor}{18.07} & 8.18 & 4.56 & 23.60 & \cellcolor{mycolor}{17.61} & 8.50 \\
       & WizardCoder-33b-V1.1  & 4.12 & 23.72 & 16.03 & 20.77 & 4.16 & 23.79 & 16.09 & 20.68 & 4.18 & 23.81 & 16.20 & 20.73 & 4.08 & 23.26 & 15.99 & 20.70 \\
       & Code Llama-7b-instruct-hf  & 4.16 & 21.10 & 16.65 & \cellcolor{mycolor}{5.55} & 4.13 & 20.70 & 16.48 & \cellcolor{mycolor}{5.42} & 4.05 & 20.67 & 16.55 & \cellcolor{mycolor}{5.31} & 4.06 & 20.99 & 16.59 & \cellcolor{mycolor}{5.50}  \\
       & Code Llama-13b-instruct-hf  & 4.22 & 19.52 & 16.46 & 7.62 & 4.20 & 19.33 & 16.22 & 7.88 & 4.25 & 19.28 & 16.42 & 7.78 & 4.19 & 19.62 & 16.45 & 8.86  \\
       & Code Llama-34b-instruct-hf  & 4.41 & 20.97 & 17.60 & 15.70 & 4.32 & 20.81 & 17.41 & 16.00 & 4.29 & 20.58 & 17.45 & 16.59 & 4.37 & 20.44 & 17.40 & 16.87 \\
       & Code Llama-70b-instruct-hf  & 4.26 & 22.37 & 16.60 & 46.33 & 4.09 & 22.18 & 16.51 & 46.31 & 4.09 & 22.39 & 16.43 & 47.83 & 4.13 & 22.14 & 16.25 & 49.08 \\
       & DeepSeek-Coder-33b-instruct & 4.66 & 24.08 & 17.38 & 17.32 & 4.67 & \cellcolor{mycolor}{24.03} & 17.38 & 17.15 & 4.62 & 24.01 & 17.42 & 16.87 & \cellcolor{mycolor}{4.61} & \cellcolor{mycolor}{23.71} & 17.28 & 17.42 \\
    \midrule
      \multicolumn{1}{c}{\multirow{8}{*}{\makecell{General \\ LLMs}}} 
       & ChatGLM2-6B  & 3.54 & 21.42 & 14.49 & 5.21 & 3.49 & 21.45 & 14.46 & 5.29 & 3.50 & 21.43 & 14.45 & 5.36 & 3.40 & 21.17 & 14.26 & 5.40 \\
       & Vicuna-7b-v1.5  & 4.47 & 23.02 & 17.64 & 4.40 & 4.39 & 22.95 & 17.58 & 4.41 & 4.41 & 22.99 & 17.68 & 4.49 & 4.36 & 22.87 & 17.44 & 4.56 \\
       & Vicuna-13b-v1.5  & 5.59 & 22.23 & 19.75 & 4.86 & 5.46 & 21.91 & 19.57 & 5.05 & 5.48 & 22.19 & 19.81 & 5.00 & 5.35 & 21.77 & 19.43 & 5.20 \\
       & Llama-2-13b-chat-hf  & 4.72 & 22.56 & 18.06 & 7.07 & 4.57 & 22.45 & 17.86 & 7.43 & 4.61 & 22.57 & 17.91 & 7.58 & 4.41 & 22.34 & 17.61 & 8.03 \\
       & Llama-2-70b-chat-hf  & 4.36 & 23.17 & 16.14 & 45.73 & 4.24 & 23.15 & 15.93 & 48.11 & 4.23 & 23.10 & 16.01 & 46.28 & 4.13 & 22.91 & 15.69 & 50.13 \\
       & Mistral-7B-Instruct-v0.2  & 6.05 & 25.65 & 20.86 & \cellcolor{mycolor}{3.51} & 6.00 & 25.21 & 20.60 & \cellcolor{mycolor}{3.51} & 5.93 & 25.69 & 20.67 & \cellcolor{mycolor}{3.59} & 5.85 & 25.38 & 20.69 & \cellcolor{mycolor}{3.59} \\
       & Mixtral-8x7B-Instruct-v0.1  & 6.41 & 26.02 & 21.04 & 12.41 & 6.27 & 25.85 & 20.93 & 11.38 & 6.36 & 26.02 & 21.08 & 11.35 & 6.27 & 25.86 & 20.90 & 11.37 \\
       & ChatGPT(gpt-3.5-turbo-16k)  & \cellcolor{mycolor}{7.69} & \cellcolor{mycolor}{29.50} & \cellcolor{mycolor}{22.09} & - & \cellcolor{mycolor}{7.57} & \cellcolor{mycolor}{29.13} & \cellcolor{mycolor}{21.94} & - & \cellcolor{mycolor}{7.46} & \cellcolor{mycolor}{29.12} & \cellcolor{mycolor}{21.92} & - & \cellcolor{mycolor}{7.40} & \cellcolor{mycolor}{29.07} & \cellcolor{mycolor}{21.76} & - \\
    \midrule
    \multicolumn{1}{c}{\multirow{2}{*}{DL-based}}
        & BinT5 \citep{bint52023saner}  & 0.00 & 2.08 & 4.69 & 0.56 & 0.00 & 2.06 & 4.60 & 0.56 & 0.00 & 2.06 & 4.65 & 0.55 & 0.00 & 2.09 & 4.72 & 0.55 \\
        & HexT5 \citep{Xiong2023ase}  & \cellcolor{mycolor}{0.10} & \cellcolor{mycolor}{6.21} & \cellcolor{mycolor}{8.44} & \cellcolor{mycolor}{0.43} & \cellcolor{mycolor}{0.10} & \cellcolor{mycolor}{6.18} & \cellcolor{mycolor}{8.47} & \cellcolor{mycolor}{0.43} & \cellcolor{mycolor}{0.09} & \cellcolor{mycolor}{6.20} & \cellcolor{mycolor}{8.38} & \cellcolor{mycolor}{0.42} & \cellcolor{mycolor}{0.12} & \cellcolor{mycolor}{6.30} & \cellcolor{mycolor}{8.49} & \cellcolor{mycolor}{0.42} \\
    \bottomrule
    \end{tabular} }
    \vspace{0ex}
  \label{tab:summarization_result_optimization} }
\end{sidewaystable*}

Unlike the function name recovery task, the performance of general domain LLMs is generally significantly better than that of code domain LLMs in binary code summary tasks. This may be attributed to the different properties of the two tasks. In the function name recovery task, the output of LLMs is usually shorter and only needs to generate a function name, which is relatively simple. In contrast, the binary code summarization task requires generating longer natural language descriptions to accurately summarize the functionality and structure of binary code, which requires the model to understand more contextual information and convert it into natural language text, which is a more complex task. General domain LLMs are better at generating longer natural language descriptions due to their inherent characteristics, while code-domian LLMs have limited capabilities in this regard.

\finding{Among all LLMs, ChatGPT performs the best in the binary code summarization task across different architectures, achieving BLEU-4 scores of 7.69, 7.38, 7.80, and 8.30 on \texttt{x64},  \texttt{x86},  \texttt{ARM}, and \texttt{MIPS}, respectively. All LLMs perform best on the \texttt{MIPS} architecture, achieving an average Rouge-L score of 18.58, which represents improvements of 5.03\%, 5.69\%, and 4.03\% compared to the other architectures. General domain LLMs perform significantly better than code domain LLMs, which is attributed to its stronger long-context understanding and summarizing capabilities.}

As observed in Table~\ref{tab:summarization_result_optimization}, similar to the results across different architectures, ChatGPT outperforms all other LLMs in the BLEU-4, METEOR, and Rouge-L metrics under various compiler optimization options. It achieves scores of 7.69, 29.50, and 22.09 under the \texttt{-O0} optimization, 7.57, 29.13, and 21.94 under the \texttt{-O1} optimization, 7.46, 29.12, and 21.92 under the \texttt{-O2} optimization, and 7.40, 29.07, and 21.76 under the \texttt{-O3} optimization options, respectively. Additionally, among code-domain LLMs, WizardCoder-15b leads in both BLEU-4 and Rouge metrics, outperforming other models in the same domain with scores of 4.72 and 18.16, 4.63 and 18.00, 4.64 and 18.07, and 4.56 and 17.61 across the different optimization options. As in the case of different architectures, CodeGen25-7b and ChatGLM2-6B still perform the worst in the binary code summarization task within their respective domains.

Similar to the function name recovery task, the performance differences across all LLMs in binary function summarization at the \texttt{-O0}, \texttt{-O1}, \texttt{-O2}, and \texttt{-O3} optimization levels are also minimal, with an average Rouge-L score of 17.68, 17.55, 17.62, and 17.42, respectively. The performance gap between \texttt{-O0} and \texttt{-O3} is only 1.49\%.

\finding{ChatGPT exhibits the best performance in the binary code summarization task across different compiler optimization settings, achieving BLEU-4 scores of 7.69, 7.57, 7.46, and 7.40 for the \texttt{-O0}, \texttt{-O1}, \texttt{-O2}, and \texttt{-O3} optimization options, respectively. All LLMs exhibit minimal performance differences across different optimization levels, with the average ROUGE-L score difference between \texttt{-O0} and \texttt{-O3} being 1.49\%.}

For the deep learning-based expert models, BinT5 \citep{bint52023saner} achieves average BLEU-4, METEOR, and Rouge-L scores of 0.00, 2.06, and 4.68, respectively, across different architectures and optimization options. HexT5 \citep{Xiong2023ase} shows slight improvement, with scores of 0.10, 6.20, and 8.43, respectively. Similar to the function name recovery task, their performance still falls significantly short of LLMs.

\finding{Similar to the previous task, existing DL-based expert models perform worse than LLMs on the binary code summarization task.}

Regarding inference time, locally deployed LLMs are generally 5-6 times longer than the function name recovery task. LLMs with 6B-7B parameters usually take 3.4 to 7.4 seconds to infer a single piece of data, 13-15B scale LLMs take 4.8 to 8.9 seconds, and 33-34B scale LLMs takes 15.4 to 21 seconds. CodeLlama-70b takes the most of 49.08 seconds among LLMs. The DL-based model also shows the advantage of inference speed, with HexT5 taking only the shortest 0.42 seconds. 

\finding{For the binary code summarization task, inference time for locally deployed LLMs is about five times that of function name recovery.}

\vspace{-3.5ex}

\subsection{\textbf{RQ3: Factors that Significantly Affect Performance \label{sec:rq3}}}

\begin{sidewaystable*}[thp]
  \centering
  \vspace{100ex}
  \caption{Performance of prompts in the form of Few-shot. The Impr. column represents the performance improvement of Few-shot compared to Zero-shot. We mark the \colorbox{mycolor}{increase} and \colorbox{gray!30}{decrease} of the metrics. (x64\_O0)}
  \renewcommand{\arraystretch}{1.38}
  \setlength{\tabcolsep}{1.2mm}{
  \scalebox{1.00}{
    \begin{tabular}{l@{\hspace{3pt}}cccccccccccccccc}
    \toprule
     \multicolumn{1}{c}{\multirow{4}{*}{\textbf{Model}}} & \multicolumn{8}{c}{\textbf{Function Name Recovery}} & \multicolumn{8}{c}{\textbf{Binary Code Summarization}} 
     \\
     \cmidrule(r){2-9}  \cmidrule(r){10-17}
     & \multicolumn{2}{c}{\textbf{\emph{Precision}}} &  \multicolumn{2}{c}{\textbf{\emph{Recall}}} &  \multicolumn{2}{c}{\textbf{\emph{F1-score}}} &  \multicolumn{2}{c}{\textbf{\emph{Time(s)}}} &  \multicolumn{2}{c}{\textbf{\emph{BLEU-4}}} &  \multicolumn{2}{c}{\textbf{\emph{METEOR}}} &  \multicolumn{2}{c}{\textbf{\emph{Rouge-L}}} &  \multicolumn{2}{c}{\textbf{\emph{Time(s)}}} \\
     \cmidrule(r){2-3}  \cmidrule(r){4-5} \cmidrule(r){6-7}  \cmidrule(r){8-9} \cmidrule(r){10-11}  \cmidrule(r){12-13} \cmidrule(r){14-15}  \cmidrule(r){16-17}
    & \textbf{Few} & \textbf{Impr.}  & \textbf{Few} & \textbf{Impr.} & \textbf{Few} & \textbf{Impr.} & \textbf{Few} & \textbf{Impr.} & \textbf{Few} & \textbf{Impr.} & \textbf{Few} & \textbf{Impr.} & \textbf{Few} & \textbf{Impr.} & \textbf{Few} & \textbf{Impr.} \\
    \midrule
        CodeGen25-7b-instruct & 17.32 & \cellcolor{mycolor}{+7.34pt} & 16.33 & \cellcolor{mycolor}{+3.71pt} & 16.04 & \cellcolor{mycolor}{+5.98pt} & 1.88 & \cellcolor{mycolor}{+0.25s} & 3.83 & \cellcolor{mycolor}{+0.27pt} & 19.88 &\cellcolor{gray!30}{-0.88pt} & 13.56 & \cellcolor{gray!30}{-0.50pt} & 11.31 & \cellcolor{mycolor}{+4.17s}  \\
        
        WizardCoder-15b-V1.0 & 29.25 & \cellcolor{mycolor}{+5.96pt} & 26.34 & \cellcolor{mycolor}{+4.23pt} & 26.64 & \cellcolor{mycolor}{+4.88pt} & 7.85 & \cellcolor{mycolor}{+6.28s} & 4.83 & \cellcolor{mycolor}{+0.11pt} & 25.07 &\cellcolor{mycolor}{+0.97pt} & 18.89 & \cellcolor{mycolor}{+0.73pt} & 10.27 & \cellcolor{mycolor}{+2.27s}  \\

        WizardCoder-33b-V1.1 & 31.76 & \cellcolor{mycolor}{+11.19pt} & 29.28 & \cellcolor{mycolor}{+8.15pt} & 29.39 & \cellcolor{mycolor}{+9.90pt} & 4.64 & \cellcolor{gray!30}{-0.37s} & 5.16 & \cellcolor{mycolor}{+1.04pt} & 26.03 &\cellcolor{mycolor}{+2.31pt} & 18.53 & \cellcolor{mycolor}{+2.50pt} & 24.06 & \cellcolor{mycolor}{+3.29s}  \\

        Code Llama-7b-instruct-hf & 28.07 & \cellcolor{mycolor}{+0.55pt} & 26.07 & \cellcolor{mycolor}{+1.01pt} & 26.19 & \cellcolor{mycolor}{+0.96pt} & 1.15 & \cellcolor{gray!30}{-0.04s} & 4.55 & \cellcolor{mycolor}{+0.39pt} & 23.83 &\cellcolor{mycolor}{+2.73pt} & 17.81 & \cellcolor{mycolor}{+1.16pt} & 6.39 & \cellcolor{mycolor}{+0.84s}  \\

        Code Llama-13b-instruct-hf & 32.08 & \cellcolor{mycolor}{+5.44pt} & 30.03 & \cellcolor{mycolor}{+5.20pt} & 29.75 & \cellcolor{mycolor}{+5.07pt} & 2.10 & \cellcolor{mycolor}{+0.38s} & 4.55 & \cellcolor{mycolor}{+0.33pt} & 23.41 &\cellcolor{mycolor}{+3.89pt} & 17.28 & \cellcolor{mycolor}{+0.82pt} & 11.37 & \cellcolor{mycolor}{+3.75s}  \\

        Code Llama-34b-instruct-hf & 31.91 & \cellcolor{mycolor}{+3.18pt} & 30.18 & \cellcolor{mycolor}{+2.85pt} & 30.08 & \cellcolor{mycolor}{+3.33pt} & 3.84 & \cellcolor{mycolor}{+0.42s} & 4.92 & \cellcolor{mycolor}{+0.51pt} & 24.45 &\cellcolor{mycolor}{+3.48pt} & 17.89 & \cellcolor{mycolor}{+0.29pt} & 24.56 & \cellcolor{mycolor}{+8.86s}  \\

        Code Llama-70b-instruct-hf & 30.92 & \cellcolor{mycolor}{+4.35pt} & 29.55 & \cellcolor{mycolor}{+4.03pt} & 29.31 & \cellcolor{mycolor}{+4.41pt} & 12.07 & \cellcolor{mycolor}{+2.02s} & 4.74 & \cellcolor{mycolor}{+0.48pt} & 25.15 &\cellcolor{mycolor}{+2.78pt} & 17.31 & \cellcolor{mycolor}{+0.71pt} & 59.01 & \cellcolor{mycolor}{+12.68s}  \\

        DeepSeek-Coder-33b-instruct & 28.97 & \cellcolor{mycolor}{+5.41pt} & 27.78 & \cellcolor{mycolor}{+3.20pt} & 27.24 & \cellcolor{mycolor}{+4.22pt} & 7.11 & \cellcolor{mycolor}{+3.12s} & 4.82 & \cellcolor{mycolor}{+0.16pt} & 25.92 &\cellcolor{mycolor}{+1.84pt} & 16.68 & \cellcolor{gray!30}{-0.70pt} & 31.30 & \cellcolor{mycolor}{+13.98s}  \\
    \midrule
        ChatGLM2-6B & 16.50 & \cellcolor{mycolor}{+2.62pt} & 14.11 & \cellcolor{mycolor}{+1.54pt} & 14.68 & \cellcolor{mycolor}{+2.02pt} & 1.30 & \cellcolor{mycolor}{+0.22s} & 3.93 & \cellcolor{mycolor}{+0.39pt} & 22.34 &\cellcolor{mycolor}{+0.92pt} & 16.27 & \cellcolor{mycolor}{+1.78pt} & 5.64 & \cellcolor{mycolor}{+0.43s}  \\

        Vicuna-7b-v1.5 & 24.05 & \cellcolor{mycolor}{+5.77pt} & 20.93 & \cellcolor{mycolor}{+2.76pt} & 21.63 & \cellcolor{mycolor}{+4.31pt} & 1.11 & \cellcolor{gray!30}{-0.02s} & 5.07 & \cellcolor{mycolor}{+0.60pt} & 22.18 &\cellcolor{gray!30}{-0.84pt} & 18.31 & \cellcolor{mycolor}{+0.93pt} & 4.54 & \cellcolor{mycolor}{+0.14s}  \\

        Vicuna-13b-v1.5 & 27.07 & \cellcolor{mycolor}{+5.83pt} & 25.09 & \cellcolor{mycolor}{+3.55pt} & 25.09 & \cellcolor{mycolor}{+4.63pt} & 1.52 & \cellcolor{mycolor}{+0.49s} & 6.16 & \cellcolor{mycolor}{+0.57pt} & 24.58 &\cellcolor{mycolor}{+2.35pt} & 20.23 & \cellcolor{mycolor}{+0.48pt} & 8.98 & \cellcolor{mycolor}{+4.12s}  \\

        Llama-2-13b-chat-hf & 24.94 & \cellcolor{mycolor}{+1.06pt} & 23.90 & \cellcolor{mycolor}{+0.62pt} & 23.06 & \cellcolor{mycolor}{+0.38pt} & 2.08 & \cellcolor{mycolor}{+0.90s} & 5.28 & \cellcolor{mycolor}{+0.56pt} & 23.34 &\cellcolor{mycolor}{+0.78pt} & 18.24 & \cellcolor{mycolor}{+0.18pt} & 10.83 & \cellcolor{mycolor}{+3.76s}  \\

        Llama-2-70b-chat-hf & 27.26 & \cellcolor{mycolor}{+3.75pt} & 24.70 & \cellcolor{mycolor}{+3.14pt} & 25.01 & \cellcolor{mycolor}{+3.29pt} & 10.59 & \cellcolor{mycolor}{+6.13s} & 5.23 & \cellcolor{mycolor}{+0.87pt} & 24.57 &\cellcolor{mycolor}{+1.40pt} & 17.35 & \cellcolor{mycolor}{+1.21pt} & 59.72 & \cellcolor{mycolor}{+13.99s}  \\

        Mistral-7B-Instruct-v0.2 & 27.59 & \cellcolor{mycolor}{+5.67pt} & 27.51 & \cellcolor{mycolor}{+3.69pt} & 26.50 & \cellcolor{mycolor}{+4.69pt} & 1.28 & \cellcolor{mycolor}{+0.08s} & 6.67 & \cellcolor{mycolor}{+0.62pt} & 27.03 &\cellcolor{mycolor}{+1.38pt} & 21.06 & \cellcolor{mycolor}{+0.20pt} & 5.59 & \cellcolor{mycolor}{+2.08s}  \\

        Mixtral-8x7B-Instruct-v0.1 & 28.39 & \cellcolor{mycolor}{+9.05pt} & 30.35 & \cellcolor{mycolor}{+4.70pt} & 28.29 & \cellcolor{mycolor}{+7.29pt} & 2.67 & \cellcolor{mycolor}{+0.02s} & 6.84 & \cellcolor{mycolor}{+0.43pt} & 27.65 &\cellcolor{mycolor}{+1.63pt} & 21.92 & \cellcolor{mycolor}{+0.88pt} & 15.44 & \cellcolor{mycolor}{+3.03s}  \\

        ChatGPT(gpt-3.5-turbo-16k) & 27.56 & \cellcolor{mycolor}{+8.16pt} & 29.11 & \cellcolor{mycolor}{+10.70pt} & 27.41 & \cellcolor{mycolor}{+9.33pt} & - & - & 7.93 & \cellcolor{mycolor}{+0.24pt} & 30.61 &\cellcolor{mycolor}{+1.11pt} & 23.16 & \cellcolor{mycolor}{+1.07pt} & - & -  \\
    \bottomrule
    \end{tabular}
    }  
  \label{tab:fewshot_result} }
\end{sidewaystable*}

We further explore the key factors affecting the performance of LLMs in this section.
As this experiment focuses on analyzing factors including the few-shot form prompts, the length of pseudo code, and the length of symbol information, the experimental environment is fixed with the \texttt{x64} architecture and the \texttt{O0} optimization option.

\subsubsection{\textbf{Few-shot prompts}}
The pre-training datasets for LLMs contain little or no binary code, which makes directly applying LLMs to binary code understanding tasks likely not to yield optimal results. In this case, few-shot prompts become a potential solution, by providing well-designed examples to LLMs, so that LLMs can learn the unique structure and syntax of binary code and quickly adapt to new tasks. Specifically, we construct two carefully designed pairs of pseudo code and ground-truth examples, and add them to the original prompts. We conduct experiments on few-shot prompts for all LLMs in the previous experiments.
The results are shown in Table~\ref{tab:fewshot_result}. 

For the function name recovery task, both code-domain LLMs and general-domain LLMs show significant improvements compared to zero-shot prompts. Among them, the code domain WizardCoder-33b and the general domain ChatGPT exhibit the most notable improvements, with Precision, Recall, and F1-score increasing by 11.19, 8.15, and 9.90 points for WizardCoder-33b, and 8.16, 10.70, and 9.33 points for ChatGPT, respectively. The code domain LLMs show an average improvement of 5.43 in Precision, 4.05 in Recall, and 4.84 in F1-score. In comparison, the performance improvement of general domain LLMs is slightly lower, with an average improvement of 5.24, 3.84, and 4.49 on Precision, Recall, and F1-score, respectively.

For the binary code summarization task, both code domain and general domain LLMs show a slight improvement compared to the function name recovery task. Among them, WizardCoder-33b exhibits the most notable improvement, with an increase of 1.04, 2.31, and 2.50 points in the BLEU-4, METEOR, and Rouge-L metrics, respectively. However, there are still some LLMs that show a decrease in METEOR and Rouge-L metrics for this task. For example, the code domain CodeGen25-7b has decreased by 0.88 and 0.50 points in METEOR and Rouge-L, respectively, while the general domain Vicuna-7b shows a decrease of 0.84 points in METEOR.  Observing their outputs, we find that the introduction of the few-shot examples increased the length of prompts, causing more test data to exceed the maximum length of window context of the model (4096 tokens) and be truncated, resulting in a decrease in performance. Overall, the code domain LLMs show an average improvement of 0.41, 2.14, and 0.63 points in the three metrics for the summarization task, respectively. In comparison, the general domain LLMs show average improvements of 0.54, 1.09, and 0.84 points.

In addition, few-shot prompts will improve inference time in most cases. However, observing the outputs of WizardCoder-15b, we find that the few-shot prompts improve the model's ability to follow instructions, reduce the output of useless information, and thus reduce the inference time. In this case, few-shot prompts not only enhance model performance but also improve inference efficiency.

\finding{When computing resources and inference time permit, few-shot prompts can be selected to improve the performance of LLMs on function name recovery and binary code summarization tasks.}

\begin{figure}
    \centering
    \begin{subfigure}[b]{0.49\linewidth}
        \centering
        \includegraphics[width=\linewidth]{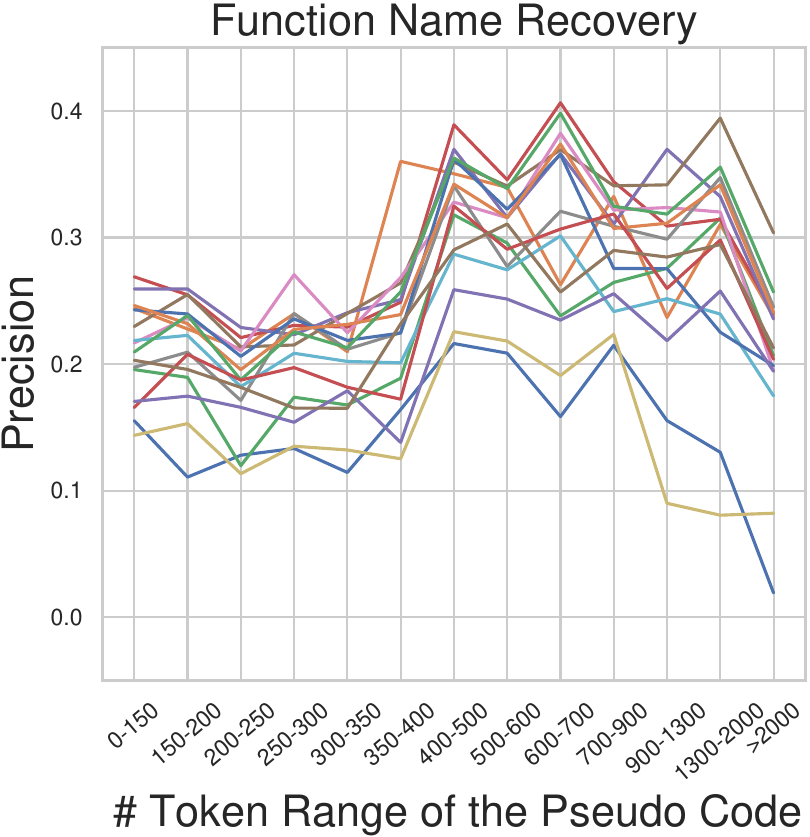}   
    \end{subfigure}
    \begin{subfigure}[b]{0.49\linewidth}
        \centering
        \includegraphics[width=\linewidth]{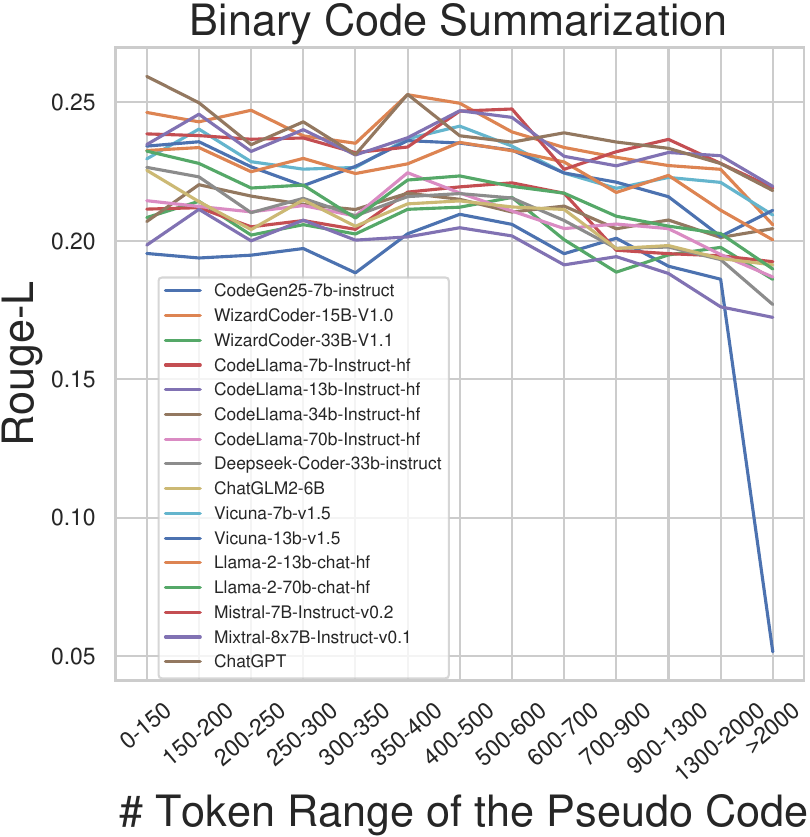}
    \end{subfigure}
    \caption{Impact of pseudo code length on performance.}
    \vspace{-1.2ex}
    \label{fig:pcodelen}
\end{figure}

\subsubsection{\textbf{Pseudo Code Length}}\label{subsec:pcodelensec}
To study the impact of pseudo code length on the performance of LLMs, we divide the length of pseudo code token according to intervals, and controll the number of data in each interval between 100 and 200 to avoid long-tail distribution of data.

As shown in Figure \ref{fig:pcodelen}, when the length of pseudo code is between 0-400 tokens, the metric of function name recovery remains at a relatively low level, as shorter pseudo code may not provide enough keywords to infer the purpose and naming intention of the function. Longer pseudo code can provide more contextual information, helping the LLMs capture semantic clues related to function names. Therefore, the metrics are relatively high between 400-2000 tokens; After exceeding 2000 tokens, the structure and logic of the code are too complex, making it difficult for the LLMs to process and integrate a large amount of information, resulting in a decrease in metrics.

For the binary code summarization task, the metrics show a slowly decreasing trend as the pseudo code length increases. As code complexity increases, LLMs find it difficult to maintain both conciseness and accuracy of summaries resulting in the generation of lengthy and unfocused summaries, thereby reducing the overall quality of the summaries.

\finding{LLMs achieve the best performance for function name recovery at moderate pseudo code length, while the performance of binary code summarization slowly decreases as pseudo code length increases.}

\begin{figure}
    \centering
    \begin{subfigure}[b]{0.49\linewidth}
        \centering
        \includegraphics[width=\linewidth]{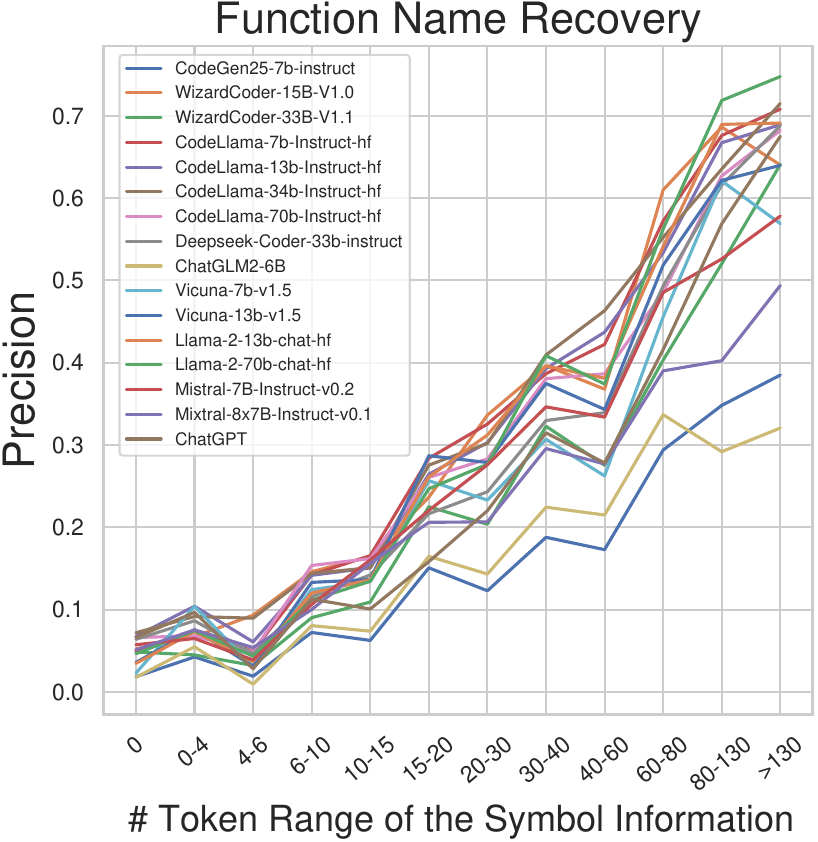}   
    \end{subfigure}
    \begin{subfigure}[b]{0.49\linewidth}
        \centering
        \includegraphics[width=\linewidth]{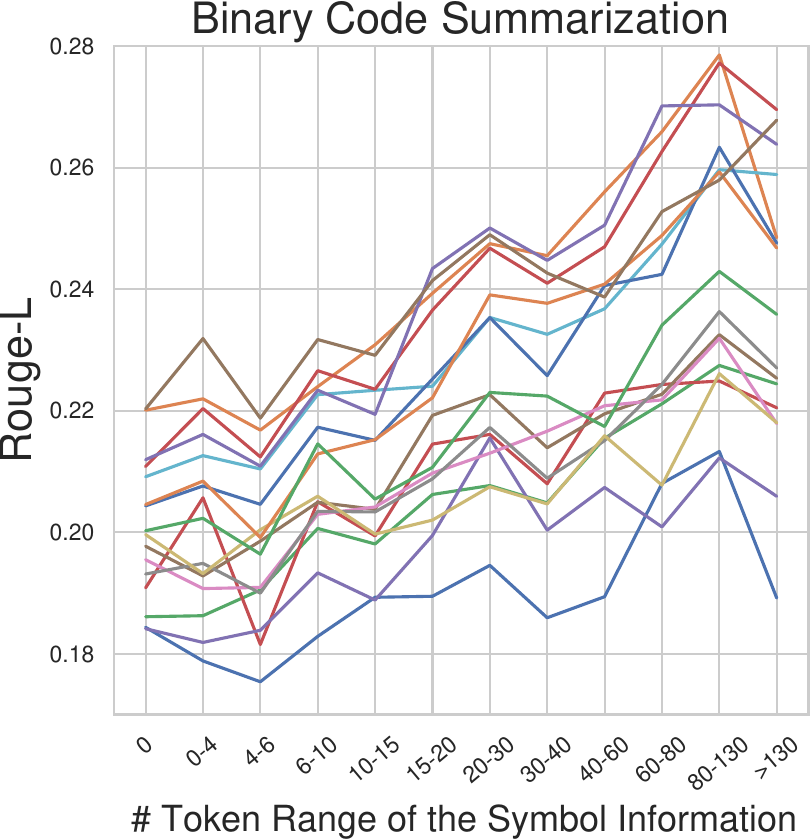}
    \end{subfigure}
    \caption{Impact of symbol information length on performance.}
    \vspace{-2ex}
    \label{fig:symbollen}
\end{figure}

\subsubsection{\textbf{Symbol Information Length}}
We define symbol information as the strings and identifiers in the pseudo code that are not stripped during the strip process, which can provide human-understandable semantic information. We also divide the length of the symbolic information token into intervals.

As shown in Figure \ref{fig:symbollen}, as the length of the symbol information token increases, the performance of both function name recovery and binary code summarization tasks increases significantly. This is due to the fact that longer symbol information provides richer semantic content and more context clues, helping LLMs understand the intent and functionality of the code. However, we found that when the symbol information token exceeds 130, the performance of most LLMs in binary code summarization tasks slightly decreases. This is because more symbol information tokens are accompanied by longer pseudo code lengths, resulting in more code being truncated due to window context limitations, affecting the completeness of LLMs summary.

\finding{The symbol information (e.g., strings and identifiers) has rich semantics and contributes significantly to LLM's understanding of binary code.} 
\vspace{-0.5ex}

\begin{figure}
    \centering
    \begin{subfigure}[b]{1\linewidth}
        \centering
        \includegraphics[width=\linewidth]{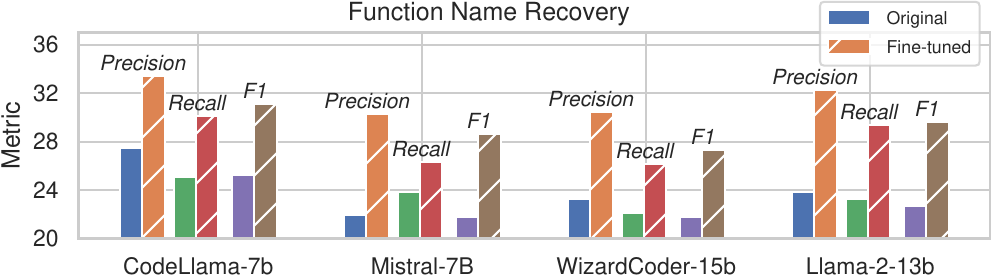}
    \end{subfigure}
    \vspace{3pt}
    \begin{subfigure}[b]{1\linewidth}
        \centering
        \includegraphics[width=\linewidth]{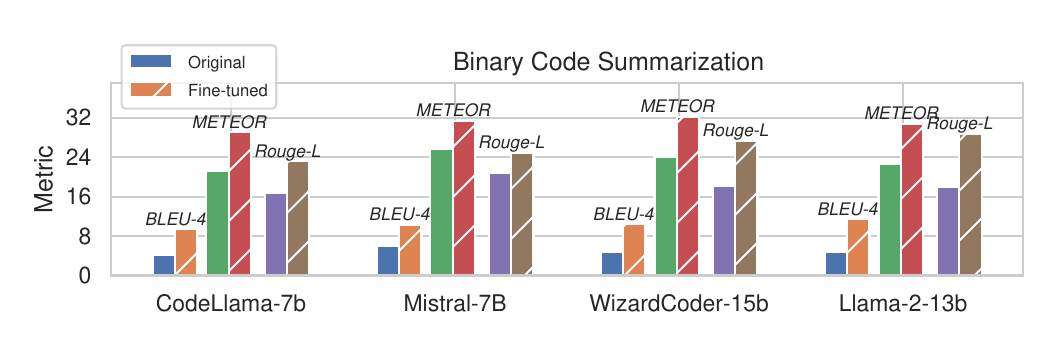}
    \end{subfigure}
    \caption{Comparison of original and fine-tuned LLMs on performance.}
    \vspace{-2ex}
    \label{fig:finetune}
\end{figure}

\subsection{\textbf{RQ4: Fine-Tuning to Enhance the Performance \label{sec:rq4}}}

As mentioned in Section \ref{subsec:finetunellm}, we build the fine-tuning dataset from the GNU repository, with each piece of data in the form of decompiled pseudo code and ground-truth pairs. Considering the computational resource limitations, we fine-tune only the 7b-15b LLMs, selecting \texttt{Codellama-7b-instruct-hf} and \texttt{WizardCoder-15b-V1.0} from the code domain, as well as \texttt{Mistral-7B-Instruct-v0.2} and \texttt{Llama-2-13b-chat-hf} from the general domain, which have performed well in previous experiments.

Figure \ref{fig:finetune} shows the performance comparison of original and fine-tuned LLMs. For the function name recovery task, the general domain \texttt{Llama-2-13b-chat-hf} shows the greatest improvement, with Precision, Recall, and F1-score increasing by 8.39, 6.06, and 6.92 points, respectively. On average, all LLMs shows improvements of 7.44, 4.42, and 6.28 points on three metrics. For the binary code summarization task, \texttt{Llama-2-13b-chat-hf} also shows the most significant improvement, with BLEU-4, METEOR, and Rouge-L metrics increasing by 6.70, 8.15, and 10.63 points, respectively. On average, all LLMs shows improvements of 5.48, 7.46, and 7.56 points. Overall, fine-tuning LLMs on downstream tasks related to binary code understanding can bring considerable performance improvements.

\finding{Introducing binary domain knowledge through fine-tuning can improve the performance of LLMs on function name recovery and summary production tasks. Among them, the general-domain \texttt{Llama-2-13b-chat-hf} demonstrates the most significant improvement in both tasks.}
\vspace{-1.5ex}

\subsection{\textbf{RQ5: Case Study on Real-World Virus Analysis \label{sec:rq5}}}

\begin{figure*}
    \centering
    \begin{subfigure}[b]{1.0\linewidth}
        \includegraphics[width=1\linewidth]{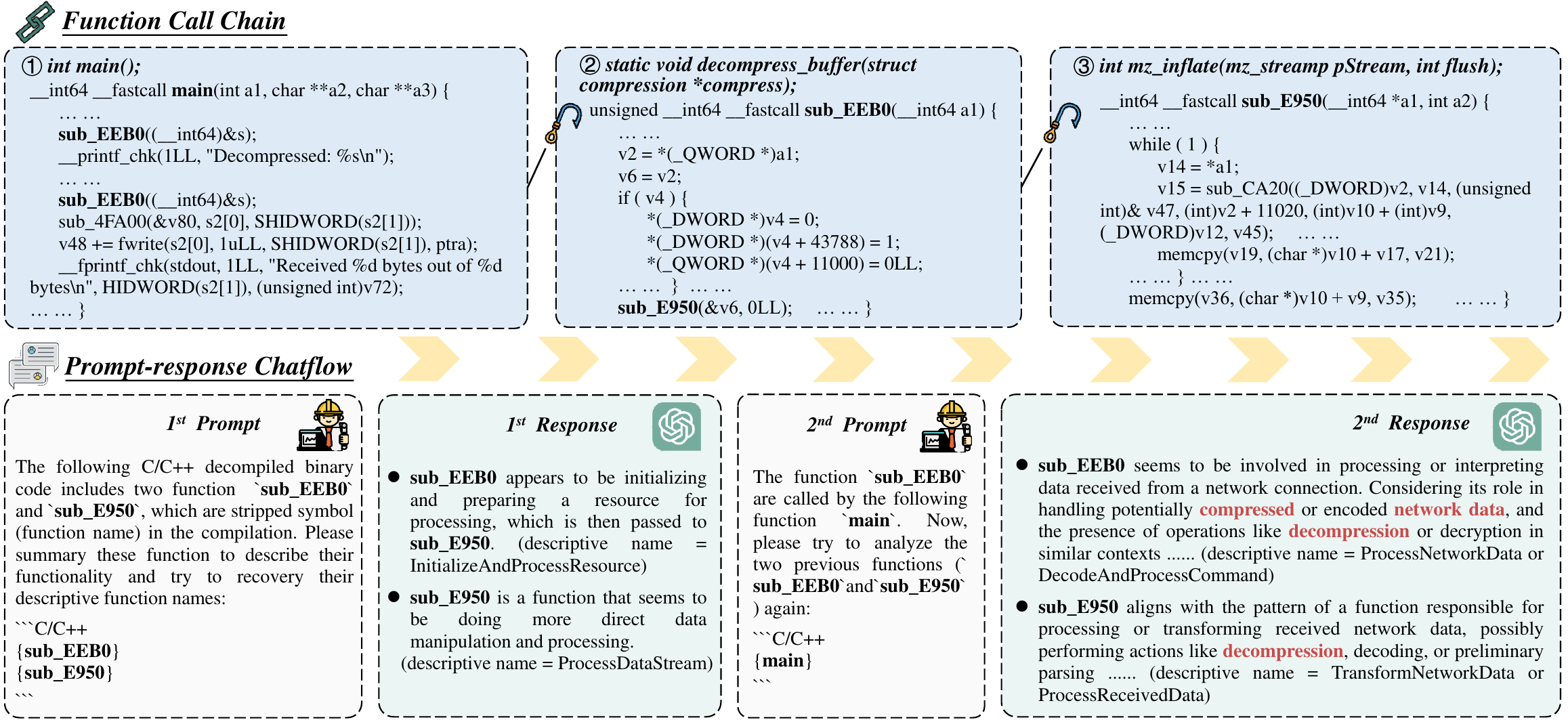}
    \end{subfigure}
    \vspace{0ex}
    \caption{An example of binary code understanding in a real-world virus with ChatGPT.}
    \vspace{-2ex}
    \label{fig:case_splinter}
\end{figure*}

We present a few case studies to show how much advanced general LLMs can assist participants in a real-world scenario. Specifically, we utilize ChatGPT \citep{chatgpt2022training} to facilitate virus analysis, including summarizing the functionality of decompiled binary functions in viruses and recovering their descriptive names. 

An open-source Linux remote access trojan named \texttt{splinter}\footnote{\url{https://github.com/tuian/splinter}} is compiled with \texttt{gcc-11.4.0} and stripped to release. In this case, Figure \ref{fig:case_splinter} has shown a partial analysis in a call chain, where the reduced pseudo code are shown in the upper part, as well as the function definitions in the source code. The lack of symbolic information makes it difficult for an analyst to understand the function \texttt{sub\_EEB0} and \texttt{sub\_E950}. We first construct the first prompt with these two functions and feed it to ChatGPT, asking for the functional summaries and descriptive names. The model gives us a primary description of the operations performed in the pseudo code without any high-level insight. We then construct the second prompt with the caller function \texttt{main} that contains a few symbol information. The second response correctly indicates that these functions are related to data decompression. At the same time, the predicted names reflect their functionality, although the predictions do not exactly match the source code. 

\begin{figure*}[htbp]
    \centering
    \includegraphics[width=1\linewidth]{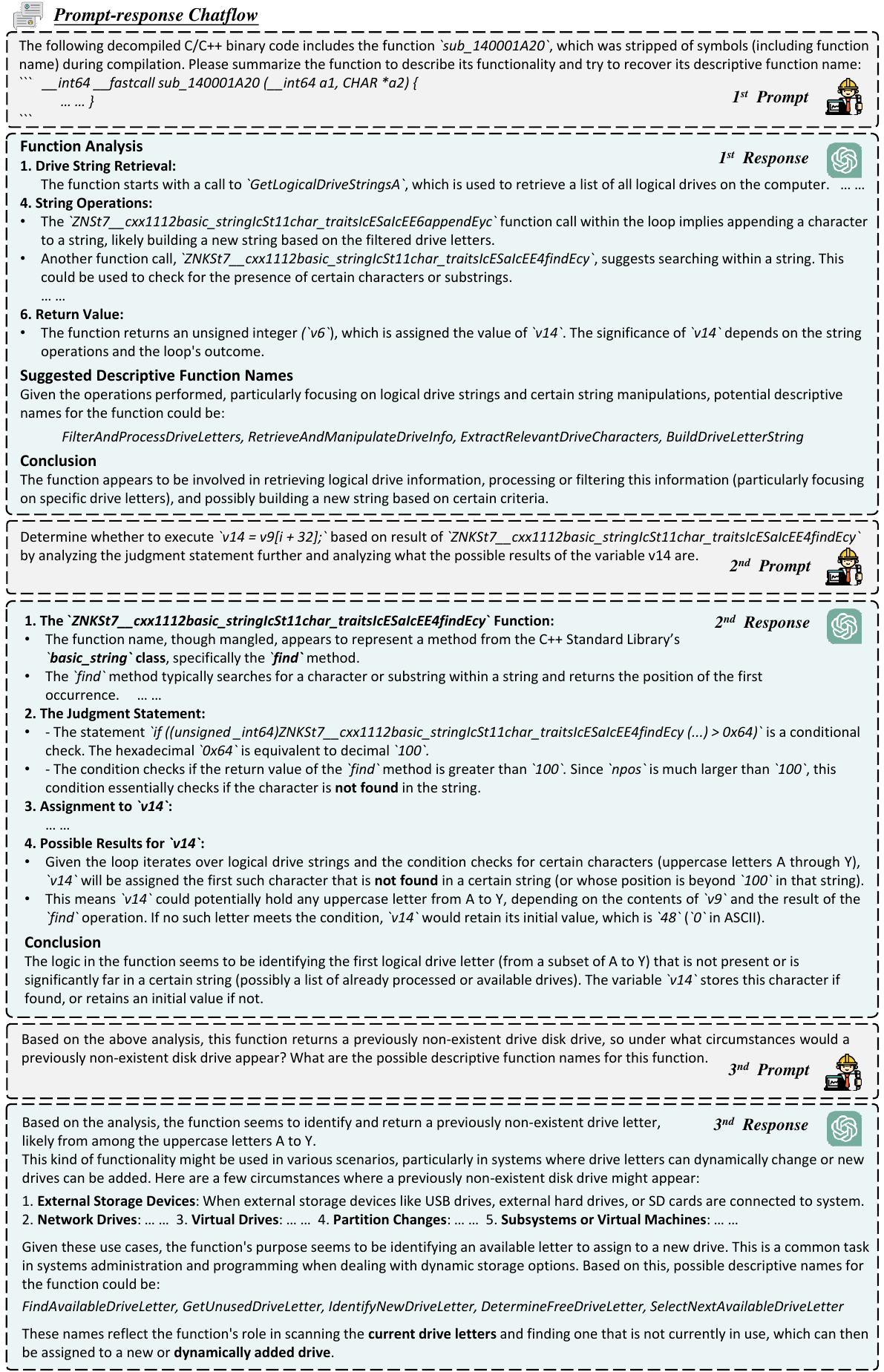}
    \caption{The chatflow of analyzing the binary function \texttt{getRemovableDisk} from \texttt{TrojanCockroach} with ChatGPT.}
    \label{fig:TrojanCockroachChatflow}
\end{figure*}

In another case, we query a function named \texttt{getRemovableDisk} in stripped binary from \texttt{TrojanCockroach}\footnote{\url{https://github.com/MinhasKamal/TrojanCockroach}} virus, which is used to get the recently inserted disk letter for further infection. In the Chatflow shown in Figure \ref{fig:TrojanCockroachChatflow}, we make several queries with ChatGPT regarding the functionality summary and function name recovery of the pseudo code \texttt{sub\_140001A20} decompiled with IDA Pro. We have omitted parts of ChatGPT's responses that deviated from our desired goals and were inconsistent with the facts. 
We find that ChatGPT provided an answer close to the fact in its first response, but it lacks further analysis of the key judgment statement. We lead it to successfully find out the callee function \texttt{find} mangled in the binary. With the help of calling context, we further prompt ChatGPT to analyze the functionality and provide high-level insights. Finally, it correctly summarizes the behavior of the function and tell us that the return value of the function is a dynamically added drive letter.

\finding{ChatGPT demonstrates the potential ability to analyze binaries in the real world. The information from the calling context will boost the predictions of LLMs.} 
\vspace{-1.2ex}

\section{Discussions \label{sec:discussions}}

Based on the experimental findings, we summarize directions for future work and limitations of the current evaluation.
\vspace{-1.2ex}
\subsection{\textbf{Future Works}}
Current LLMs have indeed shown potential in understanding binary codes, and we believe that future work can be conducted in-depth from the following aspects.

\begin{itemize}	
    \item \emph{Develop domain-specific LLM}: Binary code typically lacks annotations and rich context, making it difficult for LLMs to correctly understand the semantics and functionality of the code. Future research could focus on developing domain-specific LLMs that are pre-trained specifically for the characteristics of binary code. For example, by incorporating extensive binary domain knowledge during the pre-training phase to enhance the LLMs's grasp of code semantics and structure.
    
    \item \emph{Extend context window}: Many existing LLMs have fixed context window sizes, which are often insufficient for handling long and complex binary code. Binary code typically involves multiple functions, call relationships, and intricate control flow structures, requiring longer context windows for effective analysis. Future research should explore architectures that support longer sequence lengths, such as enhanced attention mechanisms or extended Transformer models, to better analyze complex binary code.
	
    \item \emph{Enhance processing of non-intuitive code}: Current LLMs rely on identifiers and descriptive strings in binary code to understand its functionality. However, binary code often lacks these elements, especially after being obfuscated or stripped of symbol tables.
    Future research should focus on developing new algorithms or enhancing LLM capabilities to understand binary code functionality without identifiers, using techniques such as static analysis or dynamic tracking (e.g., by integrating call chains and execution traces).
    
    \item \emph{Integrate multi-modal information}: Binary code analysis should not rely on a single source of information. By integrating multiple data sources, such as human expert annotations, assembly instructions, and dynamic execution data, LLMs' understanding of binary code can be significantly enhanced. Future research should focus on multi-modal information integration, incorporating these heterogeneous data sources into LLM inputs to provide a more comprehensive analysis of binary code.

   \item  \emph{Enhancing transfer learning capabilities}:  The behavior of binary code often varies due to compiler optimizations, target platforms (e.g., x86, x64, ARM architectures), and different operating system environments. Therefore, future LLM research should focus on enhancing models' transfer learning capabilities across different platforms. This could involve constructing training datasets that span multiple architectures and environments, enabling models to better predict function names and understand binary code in unknown platforms and environments.

    \item \emph{Robustness in analyzing obfuscated binary code}: Current binary code analysis methods struggle with obfuscated and encrypted code, especially in malware analysis. To improve LLM robustness, techniques like execution-aware code embeddings and dynamic execution tracing can help better interpret obfuscated code, recover its functionality, and enhance analysis accuracy.
    
\end{itemize}

\subsection{\textbf{Limitations}}
Although this paper provides a systematic evaluation of the performance of LLMs on binary code understanding tasks, we need to acknowledge existing limitations.
\begin{itemize}	

    \item \emph{Evaluation metrics for code summarization tasks:} Current practices are mainly based on text coherence metrics such as BLEU-4 \citep{papineni-etal-2002-bleu} and Rouge-L \citep{lin-2004-rouge}, which are originally designed for text translation tasks. However, these metrics may not be fully applicable to binary code summarization tasks. Reverse engineers typically rely on specific key terms to understand the design of a function, where text fluency is not the most critical factor. It may be beneficial to develop a new metric to better capture the essence of binary summarization.
    
    \item \emph{Binary code obfuscation:} The evaluation dataset in this paper does not consider any form of binary code obfuscation, such as encryption or compiler-based obfuscation \citep{Junod2015obf}. In fact, code obfuscation significantly alters the structure and control flow of the code, often leading to substantial changes in the form of the pseudo code, thereby increasing the difficulty of understanding and interpreting the code. Models may perform poorly when faced with obfuscated code, which could affect their applicability in real-world reverse engineering tasks. Future research could explore ways to improve the robustness of LLMs in dealing with obfuscated code, particularly in scenarios like malware analysis, where obfuscation is prevalent.

    \item \emph{Understanding function relationships:} In binary code understanding, function name recovery and code summarization are two critical tasks, but these tasks are usually handled in isolation. However, functions in binary code are usually interdependent and call each other, making the understanding of function relationships essential for comprehending the overall code. This paper focuses more on analyzing individual functions, ignoring the mutual relationships and call dependencies between different functions and the overall program structure.

\end{itemize}

\vspace{-2ex}
\section{Conclusion \label{sec:conclusion}}
Large Language Models (LLMs) have demonstrated significant potential in the field of binary code understanding. In this paper, we select two representative tasks: (1) function name recovery and (2) binary code summarization, and design an automated method to construct a benchmark for a comprehensive evaluation of LLMs' ability to understand binary code.
The research findings indicate that LLMs, particularly models such as CodeLlama, WizardCoder, and ChatGPT, have achieved impressive results in certain aspects of binary code understanding. However, the models' performance still requires improvement when dealing with more complex binary structures and unseen code samples. Additionally, we observe that LLMs' performance varies across different target architectures and compiler optimization options. In particular, LLMs perform better on the \texttt{MIPS} architecture compared to other architectures, which may be attributed to the simplified instruction set and the unified function call conventions of the \texttt{MIPS} architecture. Furthermore, code domain LLMs generally outperform general domain LLMs, as they are better equipped to handle the syntax and structure of binary code, leading to superior performance.

Therefore, we call for more research to focus on this important area of software engineering, exploring ways to improve and optimize LLMs so that they can play a more pivotal role in complex binary code analysis tasks. This will open up new possibilities and application paths in the field of understanding binary code, particularly in tasks such as reverse engineering, malware analysis, and vulnerability detection.

\begin{acknowledgements}
This work was supported in part by the Natural Science Foundation of China under Grant U20B2047, 62072421, 62002334, 62102386 and 62121002, and by Open Fund of Anhui Province Key Laboratory of Cyberspace Security Situation Awareness and Evaluation under Grant CSSAE-2021-007.
\end{acknowledgements}

\bibliographystyle{spbasic}      
\bibliography{main}


\end{document}